\documentclass[final,5p,times,twocolumn]{elsarticle}
\usepackage{amssymb}
\usepackage{amsmath}
\usepackage{blindtext}
\usepackage[utf8]{inputenc}
\usepackage[T1]{fontenc}
\usepackage{textcomp}
\usepackage{algorithm}
\usepackage{algpseudocode}
\usepackage{stmaryrd}
\usepackage{stackrel}
\usepackage{multirow}
\usepackage{titlesec}
\usepackage{enumitem}
\usepackage{mathtools}
\usepackage{fancyhdr}
\usepackage{caption}
\usepackage{tabularx}
\usepackage{graphicx}
\usepackage{amsthm}
\usepackage{hyperref} 
\usepackage{lineno}
\newtheorem{definition}{Definition}
\usepackage[english]{babel}
\usepackage[nottoc]{tocbibind}
\usepackage[draft]{fixme}
\usepackage{subcaption}
\usepackage{booktabs} 
\def\reals{\mathbb{R}}
\usepackage{url}
\newcommand{\abs}[1]{\lvert #1 \rvert}

\journal{Expert Systems with Applications}

\begin{document}

\begin{frontmatter}

\title{Multiview graph dual-attention deep learning and contrastive learning for multi-criteria recommender systems}

\author[1,2]{Saman Forouzandeh \thanks{s.forouzandeh@unsw.edu.au}} 
\author[1]{Pavel N. Krivitsky  \thanks{p.krivitsky@unsw.edu.au}}
\author[2]{Rohitash Chandra  \thanks{rohitash.chandra@unsw.edu.au}}
\affiliation[1]{Department of Statistics, School of Mathematics and Statistics, University of New South Wales, Sydney, NSW, Australia}
\affiliation[2]{Transitional Artificial Intelligence Research Group, School of Mathematics and Statistics, University of New South Wales, Sydney, Australia}

\begin{abstract}
Recommender systems leveraging deep learning models have been crucial for assisting users in selecting items aligned with their preferences and interests. However, a significant challenge persists in single-criteria recommender systems, which often overlook the diverse attributes of items that have been addressed by Multi-Criteria Recommender Systems (MCRS). Shared embedding vector for multi-criteria item ratings but have struggled to capture the nuanced relationships between users and items based on specific criteria. In this study, we present a novel representation for Multi-Criteria Recommender Systems (MCRS) based on a multi-edge bipartite graph, where each edge represents one criterion rating of items by users, and Multiview Dual Graph Attention Networks (MDGAT). Employing MDGAT is beneficial and important for adequately considering all relations between users and items, given the presence of both local (criterion-based) and global (multi-criteria) relations. Additionally, we define anchor points in each view based on similarity and employ local and global contrastive learning to distinguish between positive and negative samples across each view and the entire graph. We evaluate our method on two real-world datasets and assess its performance based on item rating predictions. The results demonstrate that our method achieves higher accuracy compared to the baseline method for predicting item ratings on the same datasets. MDGAT effectively capture the local and global impact of neighbours and the similarity between nodes. 
\end{abstract}


\begin{keyword}
Multi-Criteria Recommender System; Multiview Graph Attention Network; Contrastive Learning.
\end{keyword}

\end{frontmatter}

\section{Introduction}

The exponential growth of data across diverse domains, such as e-commerce and social media  has elevated recommender systems to the forefront of technology-driven user experiences \cite{liu2022survey}. Despite their widespread adoption, traditional recommender systems face a significant limitation since they often concentrate on a singular aspect of user preferences \cite{zheng2022survey}. This limitation becomes particularly evident when users evaluate items based on multiple criteria, a common occurrence in domains such as hospitality and entertainment, where items possess diverse features. The emergence of Multi-Criteria Recommender Systems (MCRS) \cite{shambour2021deep} has been prominent in addressing the limitations. MCRS incorporates diverse criteria such as price, quality, and brand to enhance the precision and comprehensiveness of recommendations\cite{wang2020survey,wang2022survey,mu2018survey}. These systems not only provide more precise and comprehensive recommendations but also offer enhanced decision-making and flexibility in handling diverse user preferences. They effectively address the limitations of single-criterion recommender systems when evaluating items with multiple features. In recent years, the adoption of deep learning models \cite{zhang2019deep} and graph neural networks (GNNs) \cite{wu2022graph} in recommender systems has gained substantial attention. These approaches demonstrate significant potential in augmenting recommender systems by capturing intricate relationships among users, items, and attributes and representing them as a graph structure \cite{zheng2022survey, zhang2020survey}. The embeddings generated by these methods generally exhibit acceptable accuracy in various applications.

In the landscape of MCRS, several challenges have been identified, and we discuss them as follows. Firstly, there has been scrutiny on the interpretability of deep learning models, particularly autoencoders \cite{nassar2020multi, nassar2020novel,shambour2021deep}, and the perceived lack of interpretability poses a hurdle that limits the model's capacity for their meaningful explanation for the recommendations \cite{zichen2019recommendation}. This issue is compounded by the broader challenges faced by recommender systems, especially those employing deep learning techniques, such as coping with data sparsity where not all users provide substantial ratings for a diverse set of items \cite{fan2023improving}. Furthermore, ongoing research delves into understanding the relationship between users based on various criteria within MCRS for deep learning models \cite{nassar2020multi, nassar2020novel,shambour2021deep, hassan2019evaluating, farokhi2016multi, hassan2017neural}. However, accurately predicting ratings or generating recommendations is challenging due to the necessity of considering user relationships based on specific criteria, incorporating user interests, and structuring a multi-criteria recommender system around a graph \cite{fan2023improving}. The adoption of graph neural networks (GNN) captures intricate relationships in data and enhances the predictive performance of MCRS in tasks related to rating predictions and recommendation generation \cite{singh2024comparative}. However, GNNs for MCRS introduce challenges in terms of capturing the nature of relationships between nonuniform nodes in the graph, with each node holding a distinct level of importance \cite{wang2021hgate, song2022deep, xie2020mgat}. Therefore, the adoption of a multiview approach has taken into account local and global relationships \cite{huang2023self, yao2022multi}. Locally, the significance of a node lies in its connections within each specified view, while globally, it derives from relationships across all views on the graph \cite{liu2020global, fei2023gnn}. This dual consideration is crucial for a comprehensive understanding of the intricate connections within the multiview structure. 

Multiview learning \cite{yan2021deep} has attracted growing interest and has been proven to be highly effective in practical applications by using complementary information from multiple features or modalities. The objective of multiview learning is to discover shared patterns or common feature spaces by integrating various distinct features or data sources \cite{zheng2023comprehensive}. Among the prominent approaches in multiview learning is the mapping of multiview data in a unified feature space, with the aim of maximizing the consensus among multiple views \cite{bhatt2019representation, ming2021deep}. Therefore, we can define MCRS based on the use of multiview learning, which uses information from multiple distinct perspectives or views of the data to generate recommendations. In this approach, each criterion is depicted within its own view, showcasing relationships between users and items. Within each view of multiview learning, users and items are associated with relevant relations, and users rate items based on the information presented in that particular view.

We need to employ two types of attention mechanisms when using multiview learning that various recommender systems have defined based on previous work. Graph Attention Networks (GAT) \cite{velickovic2017graph} and multiview Graph Attention Networks (MGAT)\cite{xie2020mgat} are both variants of Graph Neural Networks (GNNs) that have gained significant attention in recent years for their effectiveness in modelling relational data \cite{wu2020comprehensive}. GAT extends traditional GNNs by introducing attention mechanisms, enabling nodes to weigh the importance of their neighbours during message passing \cite{velickovic2017graph}.  GAT enables more effective and flexible information propagation throughout the graph by allowing nodes in the graph to selectively attend to their neighbours' features \cite{sun2023attention, lee2019attention}. This capability allows GAT to capture complex relationships in the graph more effectively \cite{chaudhari2021attentive}. On the other hand, MGAT extends the GAT architecture to handle multiview graph-structured data, incorporating multiple views of the graph, each representing a different type of relationship or modality \cite{xie2020mgat}. This enables MGAT to capture richer and more diverse information from the data. MGAT allows nodes in each view of the graph to selectively attend to their neighbours' features across multiple views by utilising attention mechanisms, facilitating comprehensive information aggregation and propagation \cite{xie2020mgat, yao2022multi}. Each view in MGAT unveils distinct relationships marked by sparsity and bias, necessitating a tailored approach to capture these specific characteristics. For example, Zhang et al. \cite{zhang2024hybrid} introduced a method for Point of Interest (POI) recommendation, leveraging a structural graph attention network. Sun et al. \cite{sun2024rkr}  introduced a recurrent knowledge-aware recommendation algorithm that utilises a graph attention network. Liu et al. \cite{liu2024global} introduced the Heterogeneous Graph Enhanced Category-aware Attention Network  as a solution model to address both the tasks of estimating a user's category intent and predicting items for recommendation. In RS based on MGAT, Chen et al. \cite {chen2022multi} demonstrated the effectiveness of MGAT in capturing MCRS, achieving state-of-the-art performance in travel recommendation tasks. Wang et al.  \cite {wang2023multi} utilised MGAT in music recommendations and enhanced recommendation based on its usage. Hu et al. \cite{hu2023collaborative} used MGAT to capture user preferences and views that do not like, respectively.

An attention mechanism in deep learning models \cite{vaswani2017attention} allows to focus on specific parts of the input data that are more relevant to the task at hand, effectively mimicking the human ability to pay selective attention. The attention mechanism dynamically highlights important features, thereby improving learning efficiency and prediction accuracy \cite{bahdanau2014neural, vaswani2017attention, velickovic2017graph}. Local attention directs focus towards specific subsets or regions of the data within each view, emphasising relationships within individual views \cite{zheng2020global, li2017gla}. It allows the model to prioritise and weigh the importance of features or nodes within a particular view, capturing fine-grained patterns and local dependencies present within the data \cite{liu2024global}. This localised focus enhances the model's ability to extract relevant information from each view, thereby facilitating accurate recommendations based on the unique characteristics of the data within that view. On the other hand, global attention \cite{liu2024global} enables a comprehensive evaluation of relationships across all views, providing a nuanced representation of the underlying data \cite{wang2022energy, yuan2023cross, chen2023global}. According to these definitions of local and global attention, model training follows a two-fold approach: firstly, identifying similar nodes within each view; and secondly, training on nodes that exhibit similarity across all views. Therefore, the utilisation of MDGAT in the MCRS depends on analysing nodes and dependencies locally within each view and globally across all views, incorporating each criterion.

 Inspired by the success of contrastive learning in language modelling and visual representation learning, researchers have increasingly incorporated contrastive learning into recommender systems \cite{shuai2022review}. Contrastive learning is a self-supervised learning technique that focuses on distinguishing between similar and dissimilar pairs of data points \cite{liu2021contrastive}. The core idea is to learn an embedding space where similar data points are closer together, and dissimilar ones are farther apart. This is achieved by maximising the agreement between positive pairs (similar data points) while minimising the agreement between negative pairs (dissimilar data points) \cite{chen2020simple, he2020momentum}. Contrastive learning-based recommendation systems help handle sparse data by generating robust embeddings that leverage the similarity between data points, thus mitigating the effects of sparsity \cite{yang2022knowledge}. Contrastive learning enhances the quality of the learned representations by focusing on the similarity and dissimilarity between data points, \cite{qin2024intent}. Moreover, it improves the generalisation ability of the model by ensuring that the learned embeddings are meaningful and discriminative, which is essential for making accurate recommendations \cite{yu2023xsimgcl}.

In this study, we introduce a novel framework for MCRS that utilises Multiview and Dual Graph Attention and Contrastive Learning (D-MGAC). Our framework in the initial step utilises a multiview structure on the multi-edges bipartite graph. We generate for each criterion of item one graph-based user and items based relation of them and based special criterion. We use prominent MCRS benchmark datasets to demonstrate the effectiveness of our framework, including Yahoo!Movies and BeerAdvocate. Furthermore, we extracted and curated a MCRS dataset from TripAdvisor website. The key contributions of our work are as follows: 
\begin{enumerate} 
   \item The D-MGAC model employs Multiview Dual Graph Attention (MDGAT), integrating local attention to capture dependencies within each criterion and global attention for a comprehensive evaluation across all criteria.

   \item We include local and global attention for each view and the entire graph to generate embedding using the dual attention mechanism. The dual contrastive learning model integrates both local and global similarity measures by defining anchor points in each view. In the next stage, we use contrastive learning based on one loss function including two terms to calculate it based on local and global contrastive learning. It generates recommendations based on the local and global similarity between embedding vectors through contrastive learning.
\end{enumerate}    
The paper is structured as follows. In Section 2, we provide a comprehensive review of related works. Section 3 outlines our methodology, and results are given in Section 4. Section 5 provides a discussion and finally, Section 6 concludes the study with future research directions.

\section{Related Works}

\subsection{Multi-criteria Recommender Systems}

    Adopting MCRS is crucial for platforms seeking to enhance user satisfaction, engagement, and retention. Unlike traditional Recommender Systems  that rely on single criteria like ratings or item popularity, MCRS consider diverse factors such as user demographics, item attributes, temporal dynamics, and contextual information \cite{zhang2021multi, gupta2020credibility} allowing them  to deliver more personalised and contextually relevant recommendations, catering to the varied and evolving preferences of users. The rating data in a MCRS can be represented as a three-dimensional (3D) tensor, allowing factorisation of the tensor and higher order decomposition \cite{pozo2016enhancing,papalexakis2016tensors, morise2019bayesian,hong2022sentiment}. The user similarity in recommender systems has advanced from single-criterion to multi-criteria approaches with the aid of fuzzy methods \cite{nilashi2014hybrid, kermany2017hybrid} which seeks to improve predictive accuracy and tackle the challenge of personalised recommendations amidst information overload. Multi-criteria collaborative filtering systems \cite{nassar2020novel}, such as those incorporating Adaptive Neural Fuzzy Inference Systems (ANFIS) \cite{nilashi2015multi} and Self-Organising Maps (SOMs) \cite{licen2023self}, aim to provide accurate recommendations by integrating user preferences across multiple dimensions. ANFIS combines fuzzy logic with neural networks to adaptively learn from user interactions and refine recommendations based on evolving preferences and contextual cues. ANFIS is primarily used for prediction tasks rather than classification. It combines the adaptive learning capabilities of neural networks with the interpretability of fuzzy logic to model complex relationships and make predictions based on input data \cite{zhang2021causal, jain2020tweet}. SOMs, on the other hand, leverage unsupervised learning to cluster items based on their similarity and user interactions, facilitating personalized recommendations within each cluster \cite{nilashi2014hybrid}. Together, these methods enable MCRS to effectively handle the complexity of multidimensional data and enhance recommendation accuracy.
    
    MCRS often employ two-stage methods to enhance recommendation accuracy by estimating target item ratings and learning sub-score weights \cite{adomavicius2007new, zheng2017criteria, jannach2012accuracy, nassar2020novel, li2019latent}. These methods are designed to handle the complexity of multiple criteria involved in decision-making processes, ensuring that recommendations are tailored to diverse user preferences and requirements. Subsequently, deep learning models have been pivotal in advancing MCRS. Models like sparse autoencoders, which are tailored to balance criteria \cite{kalantarnezhad2022mcrs}, and deep neural networks applied to matrix factorization \cite{sinha2022dnn}  have proven highly effective in enhancing recommendation accuracy. These approaches leverage the capacity of deep learning to handle complex interactions among multiple criteria, thereby improving the precision and relevance of recommendations. These methods leverage contextual information to improve relevance \cite{vu2022deep, krishna2023analysing} and consider multiple stakeholder preferences to boost overall performance \cite{shrivastava2023deep}. For instance, Shambour et al. \cite{shambour2021deep} introduced a deep learning algorithm for multi-criteria recommenders, employing deep autoencoders to uncover intricate user preferences. Nassar et al. \cite{nassar2020multi, nassar2020novel} integrated multi-criteria recommendation with deep learning, extracting features and utilising neural networks for correlation learning. These advancements underscore the potential of deep learning approaches in enhancing the capabilities of multi-criteria recommendation systems. Hong et al. \cite{hong2021multi} introduced two single tensor models that incorporate users (or countries), items, multi-criteria ratings, and cultural groups to simultaneously account for the inherent structure and interrelationships of these elements in recommendation systems. Zhang et al. \cite{ zhang2021multi} developed methods to improve the precision and scalability of multi-criteria recommendation systems by utilising both social connections and criteria preferences. They present a hybrid social recommendation algorithm and broaden its effectiveness with an implicit technique for inferring social relationships. Rismala et al. \cite{rismala2024personalized} presented a collaborative filtering approach based on a personalised neural network that takes multiple criteria to extract individual details from the user's rating history and to represent it as rating tendencies and user experiences. Singh et al. \cite{singh2024comparative}, examined the performance of traditional collaborative filtering, matrix factorisation, and deep matrix factorisation techniques in recommender systems using multi-criteria datasets.
    
\subsection{Recommender Systems based on GNNs}

     GNNs have gained significant attention and are increasingly being applied in recommendation systems \cite{gwadabe2022improving}. GNNs offer a powerful framework to model interaction between user items and capture complex relationships within the data using the inherent graph structures present in the recommendation scenarios,  \cite{wu2020diffnet++}. These models have demonstrated effectiveness in various recommender systems, including item recommendation, user preference modelling, and explanation of recommendations \cite{wang2019neural}. Wang et al. \cite{wang2019neural} employed GNNs to propagate embeddings in the user-item bipartite graph, explicitly capturing higher-order connections and injecting collaborative signals into the recommender system. GNNs for social recommender systems \cite{fan2019graph} considered the varying strengths of social relationships among users, modelling graph information to capture interactions and opinions within the user item graph. He et al. \cite{he2017neural} used a Multilayer feedback neural network for user-item assessment modelling, while their study \cite{he2018outer} proposed a CF and convolutional neural network approach that employed dyadic product for more meaningful connections. Zheng et al. \cite{zheng2017joint} integrated user comments and items using a convolutional neural network, and Liu et al. \cite{liu2020hybrid} utilised a multilayer perceptron (MLP) in a note-based coder. Berkani et al. \cite{berkani2022neural} proposed a hybrid recommendation approach that integrates collaborative filtering (CF) and content-based filtering (CBF) within an architecture featuring two models: generalised matrix factorization (GMF) and hybrid multilayer perceptron (HybMLP). The primary aim of their model is to mitigate challenges encountered during cold start situations in recommendation systems.
    
\subsection{Recommender Systems based on Multiview learning}

    Multiview learning in recommender systems integrates diverse data sources to enhance recommendation accuracy and robustness \cite{li2015deep}. Unlike traditional systems that rely on single sources like user-item interactions or item attributes,  Multiview learning incorporates additional perspectives such as social network data, textual descriptions, temporal dynamics, and user demographics  \cite{cheng2022modeling}.By combining these varied sources,  Multiview learning mitigates data sparsity, improves prediction accuracy, and delivers personalized recommendations tailored to individual preferences and contexts \cite{zhao2017multi, palomares2018multi}. This approach also captures intricate relationships across different data modalities, thereby enhancing the overall quality and relevance of recommendations  \cite{li2015deep}. Barkan et al. \cite{barkan2019cb2cf} introduced CB2CF, a deep neural multiview model that connects item content to their collaborative filtering  representations. Designed for Microsoft Store services, which serve approximately a billion users globally, CB2CF is a real-world algorithm. The model is applied to movie and app recommendations, demonstrating superior performance over an alternative content-based model, particularly for completely new items. In paper \cite{zheng2022multiview} introduced a multiview graph collaborative filtering network for recommendations by leveraging both homogeneous and heterogeneous signals from attribute and neighbor views. The MVGCF model combines the co-occurrence features of different attribute values with the collaborative preferences of various neighbors to learn the embedding representation of nodes. The research \cite{li2024md}, introduced a deep graph collaborative signal aggregation module designed to learn latent intention similarity representations for effective collaborative signal propagation within a few layers. Additionally, they propose a novel multiview contrastive learning module, which leverages both local and global contrastive learning views derived from the collaborative signal aggregation module.  Zhou et al. \cite{zhou2024multi} proposed a multiview social recommendation framework called MsRec, which utilizes information from various perspectives for item recommendations. Specifically, MsRec aims to explore complex inner relationships within social networks and conduct user-level preference learning uniquely for each user. Additionally, MsRec incorporates available side information, such as contextual data, demographic characteristics, and item attributes, to enhance the representation vector learning for items. Yuan et al. \cite{yuan2024attribute} proposed a model called Attribute Mining Multiview Contrastive Learning Network. It enhances the initially sparse embedding representation by extracting potential information from the native data and constructing four distinct views. The model performs cross-view contrastive learning at both local and global levels, combining the collaborative information from each view with the global structural information in a self-supervised manner, thereby eliminating reliance on supervised signals.
    
\subsection{Recommender Systems based on GAT}

    Attention mechanisms have become increasingly important in recommender systems due to their ability to capture feature importance and enhance interpretability. Utilising attention allows models to focus on the most relevant parts of the input data, which is crucial for providing accurate and personalized recommendations \cite{wang2020disenhan}. The application of attention mechanisms in recommender systems offers significant benefits, including improved performance in capturing user preferences, better handling of sparse data, and enhanced interpretability of the model's decisions \cite{zhang2024hybrid}. 
    
    Wang et al \cite{wang2023multi} introduced the Multiview Graph Attention Network   for session-based music recommendation. MEGAN employs graph neural networks and attention mechanisms to generate meaningful representations (embeddings) of music tracks and users using heterogeneous data sources. This approach enables MEGAN to capture users' mixed preferences from their listening behaviors and recommend music pieces that align with users' real-time needs. Hu et al. \cite{hu2023collaborative} proposed a Collaborative Recommendation Model based on Multi-modal multiview Attention Network to represent users using both preference and dislike perspectives. Specifically, users' past interactions are categorized into positive and negative interactions, which inform the preference and dislike views, respectively. Additionally, semantic and structural details extracted from the context are incorporated to enhance the representation of items.
    
    He et al. \cite{he2023aain} categorised fundamental feature interactions into sum-interactions and product-interactions, advancing existing methods for explicit feature interactions. Building on these theoretical insights, they introduce the Attentional Aggregative Interaction Network (AAIN), a novel model that incorporates cyclic explicit modules to capture higher-order features. Their approach employs attention mechanisms to reorganise individual features, followed by product-interactions and compression of higher-order features for output. Several advancements in modelling techniques incorporating attention mechanisms have been proposed in recent literature. For example, Xiao et al. \cite{xiao2017attentional} introduced attentional mechanisms, Zhou et al. \cite{zhou2018deep} explored attentive activation mechanisms, and Song et al. \cite{song2019autoint} innovated with self-attentive neural networks. These studies represent diverse approaches aimed at enhancing model performance through attention-based methodologies. Sequential recommendation techniques include hybrid encoders from Li et al. \cite{li2017neural} and dynamic interest capture by Zhou et al. \cite{zhou2019deep}. Wang et al. \cite{wang2021hgate} introduced HGATE for unsupervised representation learning on heterogeneous graph-structured data. Chen et al \cite{chen2022multi},  the MV-GAN (Multiview Graph Attention Network) model tailored for travel recommendation that enhanced semantic understanding of users and travel products by aggregating neighbours guided by meta-paths and fusing multiple views within a heterogeneous recommendation graph. The node-level and path-level attention networks were  designed to capture user and product representations from each view independently. To integrate diverse relationship types across views, they propose a view-level attention mechanism that aggregates node representations to derive comprehensive global representations of users and products. Lin et al. \cite{lin2024mixed} introduced a Mixed Attention Network that integrates local and global attention modules for extracting domain-specific and cross-domain information. They first proposed a local/global encoding layer to capture sequential patterns specific to each domain and across domains. Finally, they proposed a local/global prediction layer to further refine and integrate domain-specific and cross-domain interests. Wang et al. \cite{wang2024sequential} introduced a sequence recommendation model to strike a balance between users' long-term and short-term benefits, ultimately combining them into a hybrid representation for recommendation purposes.
    
    Wang et al.\cite{wang2024mmkdgat} introduced to enhance recommender system for  image active recommendation that harnesssed multi-modal information and high-order collaborative signals to aid in representation learning, while also capturing personalized user preferences in images.  Li et al.  \cite{li2024multi} introduced  multiview social recommendation for item recommendation from various perspectives  to take advantage of the complex relationships within social networks. It performs user-level preference learning for each user independently, without overlap. Chen et al. \cite{chen2023knowledge} constructed an authentic global graph derived from a multiview representation of items and sessions based on a knowledge graph to extract global item-item relationships within the knowledge view for session-based recommendation.
    
\subsection{Recommender Systems based on Contrastive Learning}

    Contrastive learning \cite{chen2020simple} is a powerful paradigm in machine learning that aims to learn representations by contrasting positive pairs (similar samples) and negative pairs (dissimilar samples) in a latent space. This approach leverages the idea that similar samples should be closer together while dissimilar ones should be farther apart, thereby facilitating the discovery of meaningful patterns and representations from data. The use of contrastive learning in recommender systems has increased, with various versions being implemented. By leveraging the strengths of contrastive learning, these systems aim to learn more robust and discriminative representations of user-item interactions \cite{shuai2022review}.  Contrastive learning helps in distinguishing between positive and negative samples more effectively, thereby enhancing the quality of recommendations \cite{liu2021contrastive}.
    
    Wei et al. \cite{wei2021contrastive}  reframed the learning of cold-start item representations from an information-theoretic perspective, aiming to maximize the mutual dependencies between item content and collaborative signals. They introduce a new objective function based on contrastive learning and develop a straightforward yet effective framework for cold-start recommendations. Chen et al. \cite{chen2022intent} introduced Intent Contrastive Learning to  integrate a latent intent variable into sequential recommendation that involves learning distribution functions of users' intents from unlabeled sequences of user behaviour through contrastive self-supervised learning.
    
    Yang et al. \cite{yang2022knowledge} introduced Knowledge-Adaptive Contrastive Learning that involves  data augmentation independently from the user-item interaction view and the knowledge graph  view, and applying contrastive learning across these two perspectives. The algorithm ensures that item representations encode information that is common across both views by incorporating a contrastive loss. Zhang et al. \cite{zhang2024recdcl} developed a dual contrastive learning recommendation framework. Zhang et al. \cite{zhang2024recdcl} introduced a dual contrastive learning recommendation framework. The first contrastive learning step promotes uniform distributions across users and items, while the second step aims to generate contrastive embeddings from output vectors. Qin et al. \cite{qin2024intent} introduced Intent Contrastive Learning  for Sequential Recommendation designed to capture users' latent intentions by segmenting a user's sequential behaviour into multiple subsequences using a dynamic sliding operation. These subsequences are then processed through an encoder to generate representations that capture the user's intentions.
    
    In this paper \cite{yang2022supervised}, proposed a learning paradigm called supervised contrastive learning based on graph convolutional neural networks. Initially, during data preprocessing, they calculate the similarity between different nodes on both the user side and the item side. When applying contrastive learning, they consider not only the augmented samples as positive samples but also a certain number of augmented samples of similar nodes as positive samples. Wei et al. \cite{wei2024multi} proposed a recommendation framework called Multi-level Cross-modal Contrastive Learning. This framework aims to construct multi-level contrastive learning to fully exploit both intra-modal and inter-modal semantic information in a self-supervised manner. They consider user interaction and semantic review as two distinct semantic modalities and devise two modal-specific contrastive learning strategies to enhance intra-modal learning. 
    
    The comparison of various methodologies using single and multi-criteria recommender systems is presented in \ref{app:comparison}.

\section{Methodology}

\subsection{Definitions}
In this section, we provide some definitions for the main concepts used in the proposed method. Table \ref{tab:Notations} represents the notation and their corresponding descriptions in a clear and concise manner for better readability.

\begin{table*}[htbp!]
\centering
\small
\begin{tabular}{|c|p{10cm}|}
\hline
\textbf{Symbol} & \textbf{Definition} \\ \hline
\(U\) and \(u\in U\) & Set of users and user index \\\hline
\(V\) and \(v\in V\) & Set of items and item index \\\hline
\(i,j \in U\cup V\) & node indices \\\hline
\(C\) & number of criteria \\\hline
\( R: U \times V \mapsto \mathbb{R}^C \) & Rating function mapping users and items to ratings across \(C\) criteria. \\ \hline
\( G = (U, V, E) \) & Bipartite graph with user set \(U\), item set \(V\), and edge set \(E\). \\ \hline
\( B \in \{0,1\}^{N \times M} \) & Incidence matrix for bipartite graph \(G\). \(B_{u,v}=1\) if \(u\) rates \(v\), 0 otherwise. \\ \hline
\( B' \in \mathbb{R}^{(N+M) \times (N+M)} \) & Extended adjacency matrix for L-BGNN. \\ \hline
\( A \in \mathbb{R}^{|V| \times |V|} \) & Adjacency matrix for relationships between nodes in a given criterion. \\ \hline
\( W \in \mathbb{R}^{F' \times F} \) & Weight matrix for the linear transformation of node features. \\ \hline
\( N_{i,k} \) & Set of neighboring nodes for node \(i\) in criterion \(c\). \\ \hline
\( \alpha_{ij,k}^h \) & Attention coefficient for the \(h\)-th attention head between nodes \(i\) and \(j\) in criterion \(c\). \\ \hline
\( W_k^h \in \mathbb{R}^{F' \times F} \) & Weight matrix for the linear transformation in the \(h\)-th attention head in criterion \(c\). \\ \hline
\( a_{j,k} \) & Feature representation of node \(j\) in criterion \(c\). \\ \hline
\( a_{i,G} \) & Global attention score for node \(i\) after softmax normalization. \\ \hline
\( \mathbf{E}_c \in \mathbb{R}^{|V| \times d} \) & Embedding matrix for criterion \(c\), where each row represents a node embedding. \\ \hline
\( \mathbf{e}_c^v \in \mathbb{R}^d \) & Embedding of node \(v\) in criterion \(c\). \\ \hline
\( \mathbf{A}_c \in \mathbb{R}^{|V| \times |V|} \) & Adjacency matrix for criterion \(c\). \\ \hline
\( \mathcal{L}_{\text{LCL}} \) & Local contrastive loss function. \\ \hline
\( \mathcal{L}_{\text{HGCL}} \) & Global contrastive loss function. \\ \hline
\end{tabular}
\caption{Table of Symbols and Definitions}
\label{tab:Notations}
\end{table*}

\begin{definition}[Multi-Criteria Recommender System]

In a MCRS, the rating function is adapted to handle different ratings \cite{adomavicius2011multi}, resulting in the following representation. Let $U$ be the set of $N$ users, $V$ the set of $M$ items, and let $C$ the number of criteria, indexed from 1 to $C$. Then, ratings are represented as a function
\begin{subequations}
\begin{equation}\label{eq:er1}
R: U\times V \mapsto \reals^C
\end{equation}
or
\begin{equation}\label{eq:er1r0}
R: U\times V \mapsto \reals\times\reals^C
\end{equation}
if an overall rating (often the average of ratings over all criteria, but sometimes a separate measurement) is also present.
\end{subequations}
We will use $R_{u, v, c}$ to denote the rating given by the user $u\in U$ for the item $v\in V$ according to the criterion $c\in {1,\dotsc,C}$ or $c\in {0,\dotsc,C}$ if an overall rating is observed.
\end{definition}

\begin{definition}[Bipartite graph]

Let $G=(U, V, E)$ be a bipartite graph, where $U$ and $V$ represent the two sets of vertices. For notational convenience, we will index them as $U=\{1,\dotsc, N\}$ and $V=\{1,\dotsc,M\}$. Then, $E\subseteq U\times V$ the set of edges in the graph, representing whether user-item pairs in which the user has rated the item. In the present contaxt of multicriteria ratings, we define $G=(U, V, E_1, \dotsc, E_C)$, with $(u,v) \in E_c$ corresponding to the user $u$' rating of the item $v$ according to criterion $c$, with an analogous formulation if an overall rating is observed.

The incidence matrix for this graph is denoted as $B \in \{0,1\}^{N \times M}$, such that $B_{u,v}=1$ if and only if $(u,v) \in E$. We may, analogously, use $B_c$ to denote the incidence matrix for criterion $c$. Where the graph is weighted, $B\in\reals^{N\times M}$, can represent the weights as well.
\end{definition}

\begin{definition}[L-BGNN Adjacency Matrix]

He et al. \cite{he2019cascade} introduced layerwise-trained Bipartite Graph Neural Networks (L-BGNN) for learning node representations in bipartite graphs, where nodes are divided into two distinct sets, typically representing users and items. They propose a new format for the adjacency matrix to enhance node representation learning:
\begin{equation} \label{eq:BGNN}
    B' = \begin{bmatrix}
0 & B \\
B^T & 0 \\
\end{bmatrix} \in \reals^{(N+M)\times(M+N)}.
\end{equation}
In this matrix, \(B\) represents the connections between nodes from the two sets. A feature matrix is used to capture the relationships between these nodes, incorporating edge weights that indicate the scores assigned by users to corresponding items based on specific criteria. To improve stability during learning, we normalize \(B'\) as follows: 
\begin{equation} \label{eq:Norm_BGNN}
B' \mapsto (D^{-1} B' + B' D^{-1})/2,
\end{equation}
where \(D\) is a diagonal matrix defined by \(D_{i,i} = \sum_{j=1}^{N+M} B'_{i,j}\). This normalization helps ensure balanced propagation of information across the network, thereby enhancing the overall performance of the L-BGNN.

\end{definition}

\begin{definition}[Multiview Network Embedding]

In the context of network embedding, we consider both single-view and multiview information networks. In a single-view network denoted $G = (U, V, E)$, network embedding techniques are applied to learn low-dimensional vector representations $h_i \in \reals^d$ for each node $i\in U\cup V$. The dimensionality of the embeddings, $d$, captures the essential information of the network, preserving its structural and semantic properties. Moving on to multiview networks, denoted as $G = (U, V, E_{1}, E_{2}, ..., E_{C})$, the edge sets in each view, denoted as $E_{1}, E_{2}, ..., E_{C}$, capture the relationships between nodes based on specific criteria. For each view $c \in \{1,\dotsc,C\}$, the set of edges $E_c \subseteq U \times V$  characterises the relationships between the source nodes and the target nodes in that particular view. Our objective in multiview network embedding is to learn global node embedding representations ${h_i} \in R^d$, with $d$ being much smaller than the total number of nodes, $\abs{U} + \abs{V}$. By leveraging the information from multiple views, we can obtain comprehensive and informative node embeddings that capture the underlying structure and interactions within the network.
\end{definition}

\subsection{Recent studies}

We provide an overview of related recent studies on recommendation systems based on deep learning and  GNNs. These studies focus on single-criteria recommendation systems. Forouzandeh et al. \cite{forouzandeh2023new} introduced a recommender system based on heterogeneous network embedding. They utilized spectral clustering within the framework of a Heterogeneous Information Network (HIN) to generate recommendations. In two papers \cite{forouzandeh2024health, forouzandeh2024uifrs}, Forouzandeh et al. presented recommendations for healthy food and user-interest-based foods. The study \cite{forouzandeh2024health} introduced a health-aware recipe recommendation system based on a heterogeneous attention network, employing node-level and semantic-level attention to identify popular and healthy recipe nodes for user recommendations. In \cite{forouzandeh2024uifrs}, they introduced a user-interest-aware food recommender system using dual heterogeneous attention, utilising node-level attention to train meta-path-based users' food interests and semantic-level attention to uncover paths with heightened semantic significance based on node relations in the graph.

In \cite{rostami2024novel}, a novel measure is presented to calculate the popularity of a user within a group, defined based on both trust networks and user centrality. This measure is used to aggregate the preferences of different users in the group. Additionally, a time-aware similarity measure is defined and utilized in deep community detection, which is employed in single-user rating prediction. In their paper, Rostmai et al. \cite{rostami2022effective} propose a food recommendation system that utilizes deep learning-based food image clustering and user community detection. This system includes components for estimating user community-food group tendencies and an explainability module based on associative rule mining. 

\begin{figure}
\centering
  \begin{minipage}[b]{0.4\textwidth}
    \centering
    \includegraphics[scale=0.15]{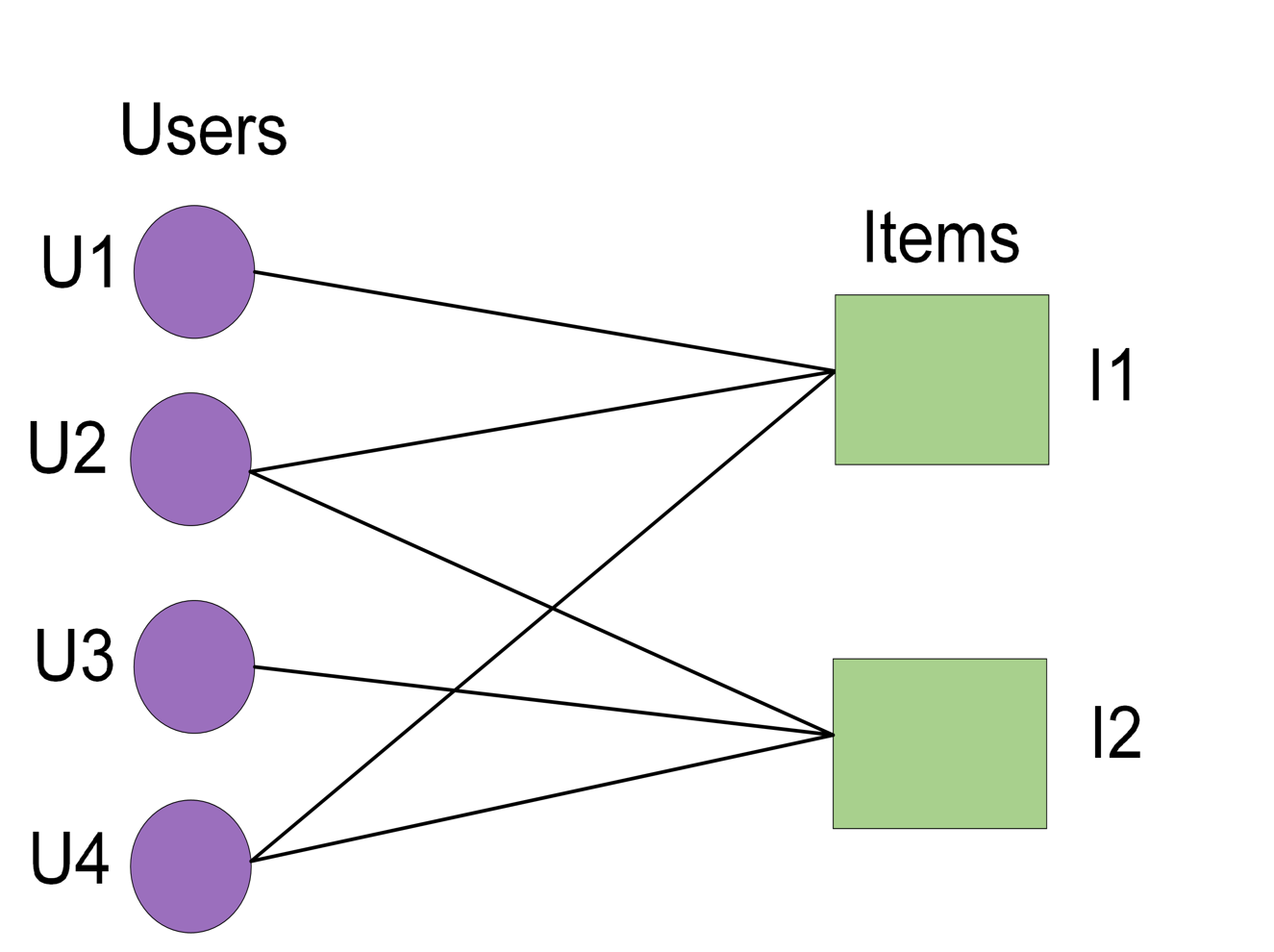}
    \subcaption{\label{subfig:Single-RS}\centering Single-Criterion recommender system}
  \end{minipage}
  \hfill
    \begin{minipage}[b]{0.4\textwidth}
    \centering
    \includegraphics[scale=0.14]{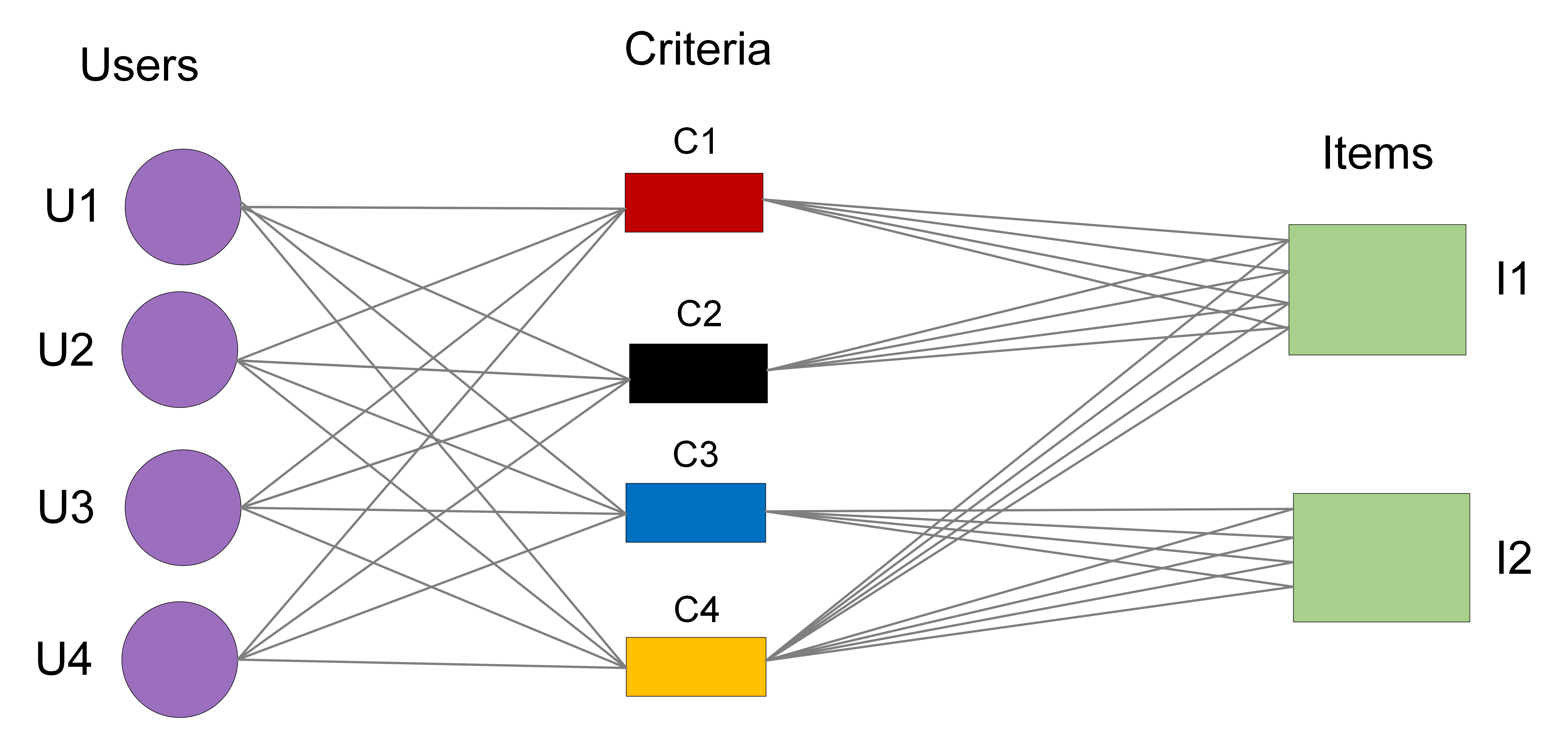}
    \subcaption{\label{subfig:Multi-RS}\centering Multi-Criteria recommender system}
  \end{minipage}
  \caption{Single and Multi-Criteria Recommendation Systems: This illustration depicts four users \( \{ u_1, u_2, u_3 \} \) and their rated items \( \{ i_1, i_2 \} \). It is important to note that single-criteria recommendation systems provide one rating per item, whereas multi-criteria systems can have multiple ratings based on the criteria \( \{ c_1, c_2, c_3, c_4 \} \).}
  \label{fig:Various_RS}
\end{figure}

However, we encountered a common issue across all of them: they focused solely on item-based ratings based on a single criterion, which only indicates how interested a user is in a particular item overall. This approach cannot fully capture the perspectives, opinions, and aspects of users who are interested in multiple features of an item. Here, in Figure~\ref{fig:Various_RS},
Panel~(\subref{subfig:Single-RS}) depicts a single-criteria recommender system where four users rate a single item, resulting in four rating scores; Panel~(\subref{subfig:Multi-RS}) depicts a multi-criteria recommender system where  the same item has received 16 rating scores due to the rating based on four criteria  from four users. For instance, consider an item such as a food dish or a hotel. Although users provide only a single rating, it may not sufficiently describe all the features of the item. For example, two users might be interested in the taste of the food or the special ingredients, while another two users might prioritise the location of the hotel. However, they may not consider factors such as the cleanliness of the hotel. Meanwhile, two other users might be more concerned about the view from the hotel. In such cases, it becomes challenging to accurately determine users' interests in the item based solely on a single criterion. This is where a  MCRS becomes valuable as it considers multiple criteria or dimensions of user preferences and generates recommendations tailored to these specific criteria. MCRS can provide more personalised and relevant recommendations to users, taking into account their diverse preferences and interests across various aspects of the items.

Previous research on MCRS \cite{nassar2020multi, nassar2020novel,shambour2021deep} utilised deep learning-based matrix structures and autoencoders to generate embeddings and analyze users' interests for generating recommendations. However, these methods often lack sufficient accuracy in the generation of recommendations \cite{fan2023improving}. This is because they do not take into account users' neighbours and fail to utilise the graph structure. Therefore, we believe there is a need to incorporate GNN structures. For example, consider Figure ~\ref{fig:Various_RS} and Section B. If two users are interested in an item based on Criteria 1, we should construct a graph and consider the neighboring nodes for them. By analysing their relationships based on the graph structure and using GNN, we can better understand their interests in the specific criteria. This approach offers advantages over traditional matrix-based analyses in accurately capturing users' interests based on special criteria.

Therefore, for this research, we opted to utilize a GNN structure with multiview and attention mechanisms, including dual attention and contrastive learning for each view and the entire graph. The proposed method aims to address the weaknesses identified in previous research by incorporating multiple views of user-item interactions. By leveraging the proposed structure introduced in the next section, our framework can learn enhanced representations, integrate diverse information sources, and effectively weigh the importance of different criteria or features. This leads to more robust and meaningful embeddings for generating recommendations.

\subsection{Unsupervised learning in recommender systems}

Unsupervised learning plays a significant role in recommender systems by enabling the identification of patterns, groupings, and structures in data without the need for labeled outputs \cite{chiu2021developing}. This type of machine learning involves algorithms that learn from data that has not been labeled, categorized, or classified. The primary goal is to uncover patterns or structures in the data that can be leveraged for making recommendations \cite{bakshi2014enhancing}. In recommender systems, unsupervised learning techniques are often employed to analyze user behaviors, item characteristics, and the relationships between items and users \cite{yassine2021intelligent}. Some methods of unsupervised learning \cite{celebi2016unsupervised, ghahramani2003unsupervised, baldi2012autoencoders, choi2019unsupervised} used in recommender systems include the following:

\begin{enumerate}
    \item \textbf{Clustering}: Clustering algorithms group similar items or users based on their features or behaviors. The underlying principle is that users with similar preferences will be clustered together, allowing the system to recommend items that are popular within that cluster.
    
    \item \textbf{Dimensionality Reduction}: Dimensionality reduction techniques reduce the number of variables under consideration, simplifying the data while retaining its essential features. This is particularly useful for high-dimensional data in recommendation tasks, utilising methods such as Principal Component Analysis (PCA) and t-Distributed Stochastic Neighbour Embedding (t-SNE).
    
    \item \textbf{Matrix Factorization}: Matrix factorization methods decompose a user-item interaction matrix into lower-dimensional matrices that capture latent factors influencing user preferences and item characteristics. Common techniques include Singular Value Decomposition (SVD) and Non-Negative Matrix Factorization (NMF).
    
    \item \textbf{Association Rule Learning}: This method discovers interesting relationships between variables in large datasets. In recommender systems, it is often employed to find item pairs that are frequently purchased or consumed together, using techniques such as the Apriori Algorithm and FP-Growth (Frequent Pattern Growth).
    
    \item \textbf{Autoencoders}:Autoencoders are a deep learning approach often combined with graph neural networks to learn efficient representations of data, typically for dimensionality reduction or feature extraction. They consist of an encoder that compresses the input into a latent space representation and a decoder that reconstructs the input from this representation. In recommender systems, autoencoders can effectively capture user preferences and item features, even from sparse interaction data.
\end{enumerate}

In this context, our proposed method, D-MGAC, harnesses the power of unsupervised learning to enhance the recommendation process. By utilising a multiview structure on a multi-edge bipartite graph, D-MGAC captures the inherent relationships between users and items across various criteria without the need for labeled data. This framework enables the identification of latent patterns and dependencies in user-item interactions, allowing the model to effectively analyze user behaviors and item characteristics.

The integration of Graph Neural Networks (GNNs) and dual attention mechanisms facilitates the extraction of meaningful embeddings that represent user preferences and item attributes. GNNs model user-item interactions in the graph structure, while local and global attention mechanisms differentiate the significance of various interactions, capturing contextual relationships without explicit labels. Additionally, our dual contrastive learning framework refines these embeddings by focusing on both local and global similarities, distinguishing between similar and dissimilar pairs. This adaptability enables D-MGAC to uncover complex patterns in user interactions and adjust to evolving user preferences and item features over time, making it a robust solution for modern recommendation tasks.

\subsection{Our Framework: Dual Multiview and  Graph Attention and Contrastive Learning}

As noted earlier, Dual-MGAT refers to the dual attention mechanism in MGAT \cite{xie2020mgat} which lays the foundation of our proposed framework known as Dual Multiview and  Graph Attention and Contrastive Learning (D-MGAC).

\begin{figure*}
\centering
\includegraphics[width=0.7 \textwidth]{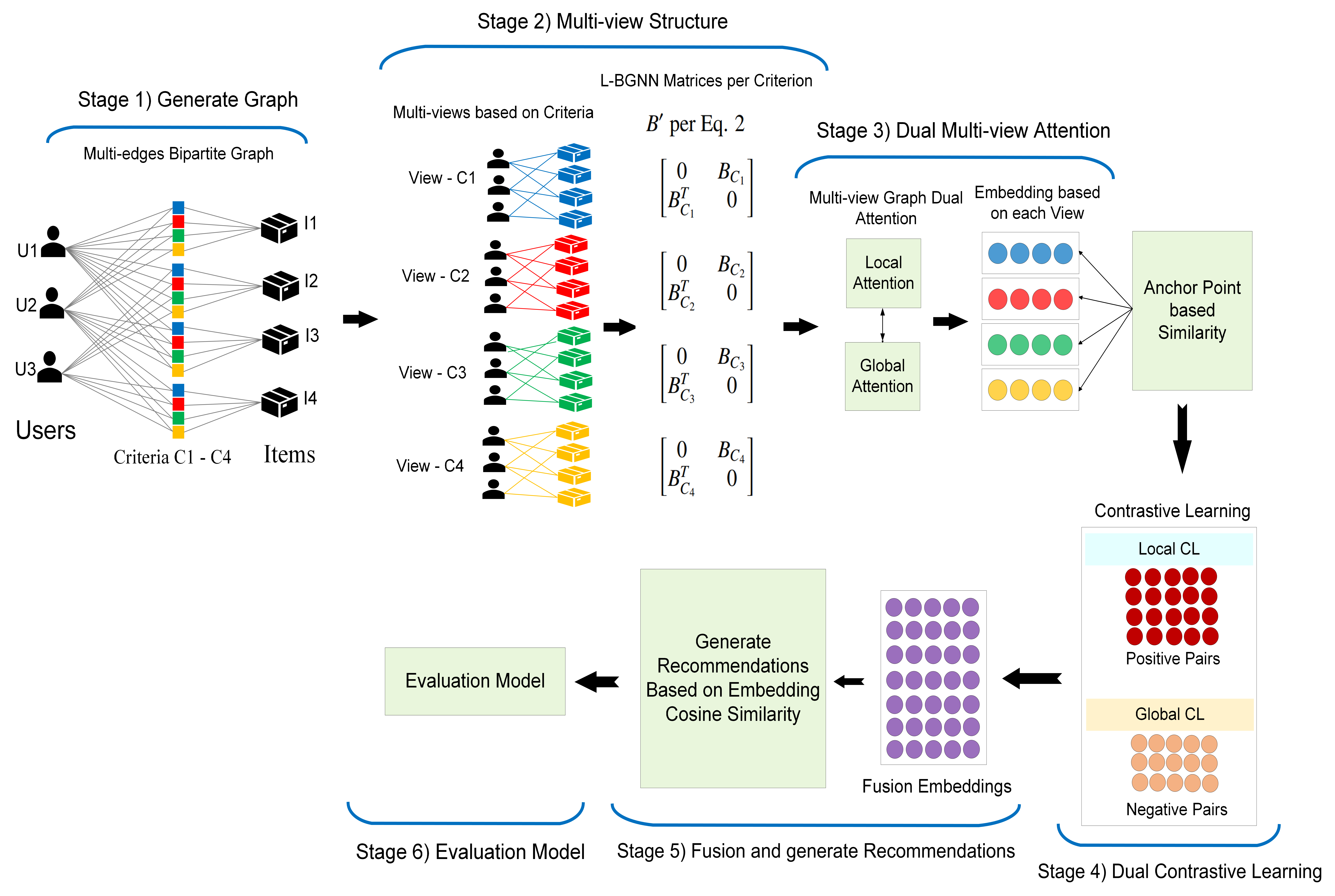}
\caption{ In Stage 1, we construct a bipartite multi-edge graph based on users \(\{ u_1, u_2, u_3 \}\), items \(\{ i_1, i_2, i_3, i_4 \}\), and criteria \(\{ c_1, c_2, c_3, c_4 \}\). Each edge in the graph represents the rating of an item based on a specific criterion. In Stage 2, we define a view for each criterion by constructing L-BGNN matrices, where each matrix captures the relationship between users and items based on the respective criterion. The matrix \(B\) represents the user-item bipartite adjacency, while \(B^T\) denotes the transposed relationship, connecting items back to users. Each \(c_i\) corresponds to a specific criterion, highlighting how user ratings are influenced by various factors. In Stage 3, we generate embeddings using dual Multi-Graph Attention Networks (MGAT), which incorporate both local and global attention mechanisms. Stage 4 focuses on defining anchor points and applying Local and Global Contrastive Learning (CL). Stage 5 emphasises fusing the embeddings to generate recommendations. Finally, in Stage 6, we evaluate the proposed model by calculating the mean of error metrics, including the mean Absolute Error (MAE) and Root Mean Square Error (RMSE), alongside the standard deviation (Std) over 30 runs. Additionally, we predict item ratings using a Support Vector Machine (SVM) model based on the training and test sets.}
\label{fig:framework}
\end{figure*}

We utilise a bipartite graph structure and dual MGAT, where each criterion is treated as a distinct view. We need to represent  the relationships between users and items as separate subgraphs.  Figure \ref{fig:framework} highlights the different stages of our method, providing a visual overview of D-MGAC. We construct an adjacency matrix using the L-BGNN to capture the proximity between nodes within each view. In addition, we consider the number of criteria rated by users for each item, resulting in the creation of an L-BGNN matrix. These matrices serve as the input for a  GAT which generates embedding vectors that incorporate user and item nodes. GAT features a dual attention mechanism that includes local and global attention. In stage 3 of our framework (Figure \ref{fig:framework}), we calculate the importance of node neighbors and perform local aggregation within each view, while simultaneously executing global aggregation across the entire graph. This results in the generation of embeddings based on the dual attention mechanism, facilitating a nuanced understanding of relationships within each view and across the entire graph. This enhances the model's capacity to capture local and global dependencies and enables the incorporation of user preferences based on multiple criteria. The local and global aggregation of attention mechanisms contribute to a comprehensive representation of the data, resulting in more accurate embeddings and improved predictive performance. In Stage 4, we employ dual contrastive learning based on local and global loss functions. Initially, for each view, we define anchor points based on similarity to determine positive and negative samples. Local contrastive learning focuses on relationships within individual views, capturing fine-grained patterns and dependencies specific to each criterion. Subsequently, global contrastive learning evaluates relationships across all views, capturing broader patterns and dependencies that span multiple criteria. Furthermore, the modular structure of our framework allows for flexibility and adaptability to various recommendation scenarios, making it a versatile solution for MCRS. 

\subsubsection{Dual-MGAT}

We present the graph attention layer used in our multiview network, coupled with an attention-based aggregation approach that enables the learning of weights for different views, leading to global node representations. Our model uses two inputs: a feature matrix \(X\) with dimensions \(|V| \times F\) and an adjacency matrix \(A\) with dimensions \(|V| \times |V|\) for each view. Here, \(F\) represents the dimension of the input features for each node, and \(A\) encodes relationships based on item ratings by users for each criterion. Node representations are defined as \(x = \{x_1, x_2, \ldots, x_{|V|}\}\), where \(x_i \in \reals^F\). The graph attention layer employs a shared linear transformation with a learnable weight matrix \(W \in \reals^{F' \times F}\) applied to each node. An attention mechanism, parameterized by a learnable weight vector \(a \in \reals^{2F'}\), is also used. The resulting node representations, denoted as \(x' = \{x'_1, x'_2, \ldots, x'_{|V|}\}\), are obtained by incorporating attention coefficients and transformed node features.

Our framework, D-MGAC integrates attention mechanisms at two levels: local and global. For local attention, we compute the attention weight matrix for each node by considering its neighbours within each view. For global attention, we determine attention weights based on the entire graph using pooling. This global attention mechanism considers nodes across all views and criteria, comprehensively evaluating relationships throughout the dataset. It provides a nuanced representation that encompasses interactions between different criteria and views. For each node \(i\) and criterion \(c\), we generate an intermediate representation, \(a_{i,c}\), capturing aggregated features from each attention head:
\begin{equation}\label{eq:LocalAtt}
a_{i,L} = \parallel^h_{h=1} \sigma\left(\sum_{j \in N_{i,c}} \alpha_{ij,c}^h \cdot W_{c}^h \cdot a_{j,c}\right)
\end{equation}
 where $a_{i,L}$ represents the output feature of node $i$ in criterion $c$, $h$ denotes the number of attention heads, $\parallel$ denotes concatenation, $\sigma$ is the activation function, $N_{i,k}$ is the set of neighboring nodes for node $i$ in criterion $c$, $\alpha_{ij,c}^h$ is the attention coefficient for the $h$-th head between nodes $i$ and $j$ in criterion $c$, $W_{c}^h$ is the weight matrix for the linear transformation within the $h$-th head in criterion $c$, and $a_{j,k}$ is the feature representation of node $j$ in criterion $c$. For global attention, we use a pooling mechanism to aggregate the information from all nodes. First, we apply a linear transformation using the weight matrix \(W_G\) to obtain \(z\), where \(z = W_G \cdot x_G\). Here, \(W_G\) is a learnable weight matrix, and \(x_G\) is the aggregated feature vector of the global attention pool. Next, we apply an activation function, such as ReLU, to the result of the linear transformation. This gives us \(z' = \max(0, z) = \max(0, W_G \cdot x_G)\). Finally, we apply the softmax function to the result of ReLU activation to obtain global attention scores. The explicit mathematical representation of the global attention computation is:
\begin{equation}\label{eq:GlobalAtt}
a_{i,G} = \frac{e^{\max(0, (W_G \cdot x_G)_i)}}{\sum_{j} e^{\max(0, (W_G \cdot x_G)_j)}}
\end{equation}
where \(a_{i,G}\) denotes the global attention score for node \(i\). This step-by-step process captures the comprehensive approach to global attention computation within our model. By combining both local and global attention mechanisms, our model captures fine-grained local patterns as well as broader global patterns, providing a comprehensive understanding of the graph structure.

\subsubsection{Contrastive Learning in MCRS}

As we explore user interactions, we observe diverse ratings across various criteria \( c \) that reflect their preferences. Each item is rated based on multiple criteria, and users may have positive ratings on some criteria and negative ratings on others. This variability poses challenges in determining user similarity and accurately classifying their preferences as positive or negative samples. Alternatively, in contrastive learning (CL), anchor points are typically selected via random sampling based on data similarity, and positive pairs are created using data augmentation techniques \cite{hua2023multimodal, xu2024contrastive}. However, the presence of noisy labels and inconsistencies in the data can complicate the process of accurately selecting anchor points and positive samples, resulting in less-than-optimal outcomes \cite{xia2019anchor}. When dealing with datasets structured around multiple criteria, the arbitrary selection of anchors becomes challenging due to the interconnected nature of these criteria and behaviors, highlighting the need to consider hierarchical relationships. To address these issues, we propose an innovative approach that eliminates the requirement for random anchor selection by leveraging similarity and hierarchical relationships.

To address these challenges, we establish anchor points within each criterion \( c \) based on their highest similarity to other nodes. These anchor points serve as positive samples, while nodes that do not exhibit similarity to any anchor point are treated as negative samples. Subsequently, we define both local and global contrastive learning strategies to measure similarity. Let \( \mathbf{E}_c \) denote the embedding matrix for criterion \( c \), where each row \( \mathbf{e}_c^v \) represents the embedding of node \( v \). \( \mathbf{A}_c \) represents the adjacency matrix for criterion \( c \), and \( N_c(v) \) denotes the set of neighbors of node \( v \) in criterion \( c \). For each node \( v \) in criterion \( c \), we calculate the average similarity:
\begin{equation}\label{eq:AnchorP}
S_c(v) = \frac{1}{|N_c(v)|} \sum_{u \in N_c(v)} \text{Sim}(\mathbf{e}_c^v, \mathbf{e}_c^u)
\end{equation}
Here, \( S_c(v) \) denotes the average cosine similarity between the embedding of node \( v \) and its neighboring nodes' embeddings in criterion \( c \). The anchor node \( v_{A} \) in criterion \( c \) is identified as the node with the highest average neighborhood similarity:
\begin{equation}\label{eq:AVEAnchor}
v_{A} = \arg\max_{v \in V} S_c(v)
\end{equation}
where \( V \) represents the set of all nodes. This approach ensures that anchor points are selected based on their local neighborhood similarity, facilitating effective local and global contrastive learning. This anchor node then serves as the reference point for subsequent positive and negative sampling.

Next, we define the loss function for local contrastive learning. Local contrastive learning focuses on maximizing the similarity between an anchor node and its positive sample while minimizing the similarity with negative samples within the same criterion. The softmax function is applied over the positive and negative similarities. For each node \( i \) in criterion \(c \), where \( e^i_c \) represents the embedding of node \( i \) in criterion \( c \), \( e^i_{c'} \) denotes the embedding of node \( i \) in another criterion \( c' \), and \( e^{\text{neg}}_{c'} \) stands for the embedding of a negative sample in criterion \( c' \), the local contrastive loss \( \mathcal{L}_{\text{LCL}} \) is formulated as:
\begin{equation}\label{eq:LocalCL}
\mathcal{L}_{\text{LCL}} = - \sum_{i=1}^{N} \sum_{c \neq c'} \log \frac{\exp\left(\frac{\text{sim}(e^i_c, e^i_{c'})}{\tau}\right)}{\exp\left(\frac{\text{sim}(e^i_c, e^i_{c'})}{\tau}\right) + \sum_{\text{neg}} \exp\left(\frac{\text{sim}(e^i_c, e^{\text{neg}}_{c'})}{\tau}\right)}
\end{equation}
Here, \( \mathcal{L}_{\text{LCL}} \) aims to maximize the similarity between embeddings \( e^i_c \) and \( e^i_{c'} \) of the same node \( i \) in different criteria \( c \) and \( c' \), while minimizing the similarity between \( e^i_c \) and negative samples \( e^{\text{neg}}_{c'} \). The temperature parameter \( \tau \) adjusts the scale of the similarity measurements.

Global contrastive learning enhances the similarity between global embeddings, which are computed as the mean of node embeddings across different criteria, while minimizing similarity with negative global embeddings. For each pair of criteria \( c \) and \( c' \), where \( g_c \) represents the global embedding (mean of node embeddings) for criterion \( c \), \( g_{c'} \) denotes the global embedding for another criterion \( c' \), and \( g_{\text{neg}} \) stands for the global embedding of a negative sample criterion, the global contrastive loss \( \mathcal{L}_{\text{HGCL}} \) is expressed as:
\begin{equation}\label{eq:GlobalCL}
\mathcal{L}_{\text{HGCL}} = - \sum_{c \neq c'} \log \frac{\exp\left(\frac{\text{sim}(g_c, g_{c'})}{\tau}\right)}{\exp\left(\frac{\text{sim}(g_c, g_{c'})}{\tau}\right) + \sum_{\text{neg}} \exp\left(\frac{\text{sim}(g_c, g_{\text{neg}})}{\tau}\right)}
\end{equation}
Here, \( \text{sim}(\cdot, \cdot) \) denotes the cosine similarity function, and \( \tau \) is the temperature parameter. \( \mathcal{L}_{\text{HGCL}} \) encourages the global embeddings \( g_c \) and \( g_{c'} \) from different criteria to be more similar to each other compared to negative global embeddings \( g_{\text{neg}} \). The temperature parameter \( \tau \) adjusts the sensitivity of the similarity measurements within the softmax function. 

The total loss function, \( \mathcal{L}_{\text{total}} \), is a combination of the local contrastive loss \( \mathcal{L}_{\text{LCL}} \), the global contrastive loss \( \mathcal{L}_{\text{HGCL}} \), and \( L2 \) regularization:
\begin{equation}\label{eq:TotalLoss}
\mathcal{L}_{\text{total}} = \alpha \cdot \mathcal{L}_{\text{LCL}} + \beta \cdot \mathcal{L}_{\text{HGCL}} + \lambda \cdot \| \mathbf{W} \|_2^2
\end{equation}
Here, \( \mathcal{L}_{\text{total}} \) represents the total loss function, which is composed of three terms weighted by hyperparameters: \( \alpha \), \( \beta \), and \( \lambda \). \( \alpha \) controls the weight of the local contrastive loss \( \mathcal{L}_{\text{LCL}} \), \( \beta \) determines the importance of the global contrastive loss \( \mathcal{L}_{\text{HGCL}} \), and \( \lambda \) regulates the impact of the \( L2 \) regularization term \( \| \mathbf{W} \|_2^2 \), which helps prevent overfitting by penalizing large parameter values.

In the next step, after obtaining the embedding vector for each criterion, which includes user nodes and items, we concatenate the obtained vectors and arrange them in a matrix. The concatenation operation for fusion can be represented as:
\begin{equation}\label{eq:Fusion}
F_{i} = \oplus_{c=1}^{c} E_{i,f}
\end{equation}
where \( F_{i} \) denotes the fused embedding vector for the \( i \)-th node, and \(C \) represents the number of criteria. The embedding vector of the \( i \)-th node in the \( c \)-th criterion is denoted as \( E_{i,f} \), and \( \oplus \) represents the concatenation operation.

\subsubsection{Collaborative Filtering Recommendation Model}

Collaborative filtering is a widely-used approach in recommendation systems that utilizes user preferences and behaviors to generate personalized recommendations. The recommendation model described here simplifies collaborative filtering by relying on cosine similarity between user embeddings. In this model, users are represented as embeddings in a latent space that captures underlying patterns in their preferences. These embeddings are derived from a multiview graph attention model. Cosine similarity is employed to measure the similarity between users using the following formula:
\begin{equation}\label{eq:Cosine}
\text{Sim}(u_i, u_j) = \frac{ \sum_{c=1}^{C} u_i^{c} \cdot u_j^{c}}{\| u_i \| \cdot \| u_j \|}
\end{equation}
Here, \( u_i \) and \( u_j \) denote the embeddings of users \( i \) and \( j \), respectively, in a \( d \)-dimensional latent space. The variable \( d \) represents the dimensionality of the embedding space. The numerator \( \sum_{k=1}^{d} u_i^{c} \cdot u_j^{c} \) calculates the dot product of the user embeddings, while the denominator \( \| u_i \| \cdot \| u_j \| \) denotes the product of the Euclidean norms (or magnitudes) of the embeddings \( u_i \) and \( u_j \). This similarity measure effectively captures the closeness between the preferences of users \( i \) and \( j \) in the latent space.The general function of D-MGAC is represented in Algorithm 1.

\begin{algorithm} 
\small
\caption{D-MGAC: MCRS based on the dual MGAT and contrastive learning}
\label{alg:D-MGAC}
\begin{algorithmic}[1]
\Statex \textbf{Input}:
\Statex Bipartite graph comprising nodes representing users and items, connected by multiple weighted edges.
\Statex L-BGNN matrix for each criterion-based view.

\Statex \textbf{Output}:
\Statex Embedding vectors generated by the Multiview Dual Graph Attention Network model.
\Statex Fusion embedding obtained from multiview embedding vectors.

\For{$m = 1$ to $\text{N}$}
    \State Define the L-BGNN matrix for each view based on each criterion.
    \State Calculate the local attention-based multi-head aggregated features \( (a_{i,k}) \), as per Eq. (\ref{eq:LocalAtt}).
    \State Calculate the global attention \( (a_{i}) \), as per Eq. (\ref{eq:GlobalAtt}).
    \State Define anchor points based on each criterion, \( v_{A} \), as shown in Eqs. (\ref{eq:AnchorP}) and (\ref{eq:AVEAnchor}).
    \State Define the local contrastive learning \( \mathcal{L}_{\text{LCL}} \), as per Eq. (\ref{eq:LocalCL}).
    \State Define the global contrastive learning \( \mathcal{L}_{\text{HGCL}} \), as per Eq. (\ref{eq:GlobalCL}).
    \State Calculate the total loss function \( (\mathcal{L}_{\text{total}}) \), as per Eq. (\ref{eq:TotalLoss}).
\EndFor

\State Obtain fusion embedding vectors from various criteria \( (F_{i}) \), as per Eq. (\ref{eq:Fusion}).
\State Generate recommendations.
\end{algorithmic}
\end{algorithm}

\section{Results}

\begin{figure}
\centering
\includegraphics[width=0.38 \textwidth]{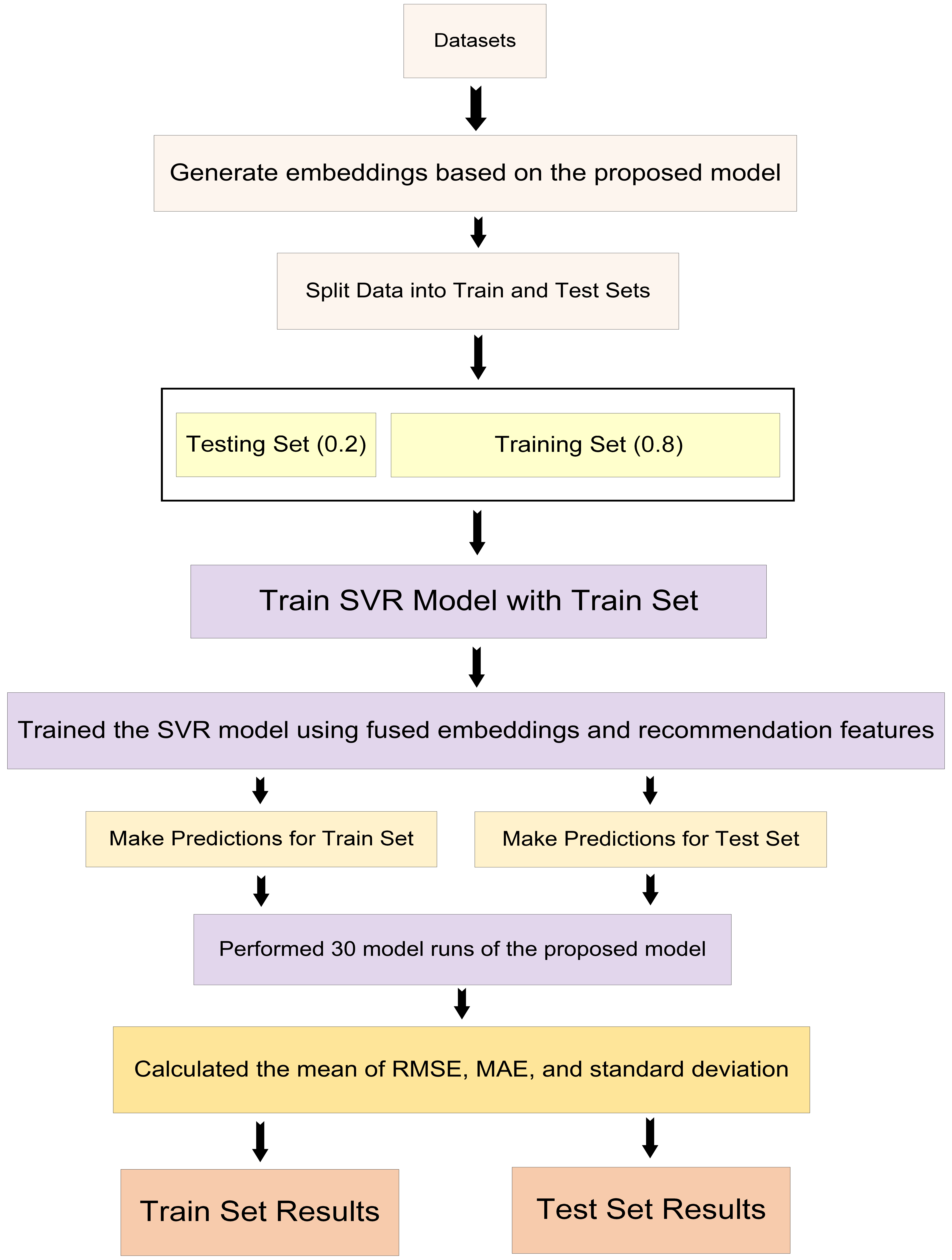}
\caption{ Evaluating the proposed model: first, the dataset is prepared and the embedding (for the whole dataset) is computed. Then, split the data into training and testing sets. Next, embeddings are extracted, and the SVR model is trained on the training set, followed by making predictions for both the training and testing sets. The model's performance is then evaluated using metrics such as RMSE, MAE, and standard deviation, calculated over 30 runs.}
\label{fig:Evaluation_Chart}
\end{figure}

This section presents an overview of the evaluation process employed in this research to assess the proposed method. The evaluation encompasses two datasets: Yahoo!Movies and BeerAdvocate. A comparative analysis is conducted between our approach and prior methodologies applied to these datasets. Each facet of the evaluation process will be introduced and elucidated in the subsequent subsections. This section presents an overview of the evaluation process employed in this research to assess the proposed method. The evaluation encompasses two datasets: Yahoo!Movies and BeerAdvocate. A comparative analysis is conducted between our approach and prior methodologies applied to these datasets. Each facet of the evaluation process will be introduced and elucidated in the subsequent subsections. The diagram for the evaluation section is shown in Figure \ref{fig:Evaluation_Chart}.

\subsection{Datasets}

The Yahoo!Movies dataset \footnote{\url{www.movies.yahoo.com}}, consists of multi-criteria movie ratings. Each user assigns an overall rating to movies, along with individual ratings for four specific criteria: story, acting, direction, and visuals. The dataset is available via GitHub \cite{nassar2020multi, nassar2020novel}, which includes 6,078 users, 976 films, and a total of 758,405 ratings. As outlined in the paper by Fan et al. \cite{fan2023improving}, we converted the scores to a range of 1 to 5 and evaluated our framework for comparison with baseline methods.

The BeerAdvocate \footnote{\url{www.beeradvocate.com}} enables users to rate four attributes (Aroma, Appearance, Palate, and Taste) of beer. After removing inactive users, the preprocessed dataset comprises 8,831 users, 2,698 items, and 3,880,359 ratings.  BeerAdvocate employs a 1–5 scaling system for both overall and multi-criteria ratings. 

We extracted and curated the TripAdvisor dataset \footnote{\url{www.tripadvisor.com}} due to several inherent challenges that significantly impacted its applicability to our study. Despite its use in various MCRS studies \cite{shambour2021deep, nassar2020multi, nassar2020novel}, we found that including this dataset could introduce substantial difficulties. To address these issues and ensure the reproducibility of our experiments, we used Zenodo to publish the extracted data instance \footnote{\url{https://sandbox.zenodo.org/records/109408}}. This platform allows us to maintain a history of dataset versions, which is crucial given the frequent changes and updates to the original data. By curating the dataset on Zenodo, we can provide a stable and consistent version for researchers, minimizing the challenges posed by the dataset’s inherent noise and sparsity. The statistics for the three datasets are shown in Table ~\ref{tab:comparison}.

\begin{table*}
\centering
\begin{tabular}{lccc}
\toprule
Statistic & \multicolumn{1}{c}{BeerAdvocate} & \multicolumn{1}{c}{Yahoo!Movies} & \multicolumn{1}{c}{Tripadvisor} \\
\midrule
Average Reviews per User & 81.531 & 10.226 & 1.389 \\
Average Reviews per Item & 266.864 & 63.684 & 15.721 \\
Sparsity & 0.970 & 0.990 & 0.999 \\
Average Number of Criteria & 4.000 & 4.000 & 7.000 \\
Variance in Criteria Ratings & 0.657 & 0.875 & 1.555 \\
\bottomrule
\end{tabular}
\caption{ \centering Comparison of statistics across datasets, highlighting why the TripAdvisor dataset is not ideal for analysis. The TripAdvisor dataset exhibits extreme sparsity (0.999), with users reviewing far fewer items (1.389 reviews per user) compared to the BeerAdvocate (81.531) and Yahoo!Movies (10.226) datasets. Additionally, TripAdvisor's higher variance in criteria ratings (1.555) suggests significant inconsistency in user evaluations across multiple criteria, complicating model performance. These characteristics indicate that TripAdvisor's data is noisier and less balanced than the other datasets, making it less suitable for standard recommendation models.}
\label{tab:comparison}
\end{table*}

As you indicated, the TripAdvisor dataset has the worst statistics compared to the other datasets. Firstly, it is characterised by considerable noise and sparsity, as evidenced by the high sparsity value of 0.999. This indicates that many users have reviewed only a limited number of items, resulting in a large proportion of missing data. This sparsity introduces challenges in maintaining reliable and consistent data, which could adversely affect the performance and validity of our model. 

Furthermore, the dataset features highly heterogeneous multi-criteria ratings across diverse travel-related categories such as hotels, restaurants, and attractions. This heterogeneity complicates the alignment and comparison with other datasets that focus on more uniform items, like movies and beers. For example, the average number of criteria ratings per item is notably high (7.000) in the TripAdvisor dataset, reflecting a complex and varied set of evaluation criteria. The variability in the criteria ratings, with a variance of 1.555, suggests significant differences in user assessments, adding another layer of complexity to the data. This variability makes it challenging to apply a consistent evaluation framework across different types of items. Additionally, the lack of standardised criteria across the various categories in the TripAdvisor dataset poses a challenge in applying a uniform evaluation framework. This inconsistency can affect the comparability and overall robustness of experiments conducted using this dataset.

Given these factors, the TripAdvisor dataset's characteristics—particularly its noise, sparsity, and heterogeneous criteria—present significant challenges. To ensure the robustness and relevance of our experiments, we have opted to exclude this dataset from our analysis and focus on datasets with more uniform and standardised criteria that better align with the objectives of our study. This approach will facilitate more consistent and reliable evaluations, ultimately enhancing the validity of our findings.

\subsection{Evaluation}

The evaluation of our model focuses on assessing the accuracy of rating predictions and, consequently, the quality of the recommendations it generates. To achieve this, we utilise two key metrics: Mean Absolute Error (MAE) and Root Mean Squared Error (RMSE). These metrics provide a comprehensive measure of how well the model's predictions align with actual user ratings. The Mean Absolute Error (MAE) is defined as:
\begin{equation} \label{eq:MAE}
\text{MAE} = \frac{1}{N}\sum_{u,i}^N |P_{u,i} - r_{u,i}|
\end{equation}
where \(N\) represents the total number of ratings, \(P_{u,i}\) is the predicted rating given by the model for user \(u\) and item \(i\), and \(r_{u,i}\) is the actual rating provided by the user for that item. MAE measures the average magnitude of the errors in a set of predictions, without considering their direction. It provides a clear indication of the average deviation of the model's predictions from the actual ratings, helping us understand how close the predicted ratings are to the real ratings. The Root Mean Squared Error (RMSE) is defined as:
\begin{equation} \label{eq:RMSE}
\text{RMSE} = \sqrt{\frac{1}{N}\sum_{u,i}^N (P_{u,i} - r_{u,i})^2}
\end{equation}
Similarly, \(N\) is the total number of ratings, \(P_{u,i}\) denotes the predicted rating for user \(u\) and item \(i\), and \(r_{u,i}\) is the actual rating. RMSE calculates the square root of the average squared differences between predicted and actual ratings. Unlike MAE, RMSE penalises larger errors more severely due to the squaring of differences. This metric helps in assessing the variance of the errors and is particularly useful in understanding how much the predictions deviate from the actual values in a squared sense. Together, MAE and RMSE offer a robust evaluation framework for rating prediction accuracy in recommender systems. MAE provides an average error magnitude, giving an intuitive sense of model performance, while RMSE highlights the impact of larger prediction errors.

\subsection{Baseline methods}

The following models and papers are relevant to baseline algorithms and belong to multi-criteria recommendation systems based on deep learning methods. Therefore,  we compare our method (D-MGAC) with the results of several previous methods from these papers, as outlined below:

\begin{enumerate}

\item UserKNN employs the standard user-based collaborative filtering technique using K-Nearest Neighbours (KNN) \cite{peterson2009k}. We select the top 100 users are selected for the target user employing Pearson correlation to measure similarities between pairs of users \cite{fan2023improving}.

\item MultiUserKNN extends UserKNN by calculating user similarity for each criterion and averaging these similarities across all criteria, using the 100 nearest neighbors to determine the final user similarity \cite{fan2023improving}.

\item \ Biased Matrix Factorization (BMF) is a collaborative filtering technique that decomposes the user-item rating matrix into latent factors to predict missing ratings. This method incorporates both user and item biases into the matrix factorization process to account for inherent biases in the data \cite{paterek2007improving}. In BMF, only the overall ratings are used to train the model parameters. Consequently, for predictions, BMF relies solely on the overall ratings and does not consider ratings for item criteria.

\item  Multilinear Singular Value Decomposition (MSVD) integrates explicit and implicit relations among users, items, and criteria. The approximation tensor is obtained by preserving the largest k-model singular values \cite{li2008improving}.

\item   Multiple Linear Regression (MLR) explores the relationship between a user's multi-criteria ratings and their overall rating to predict individual overall ratings \cite{fuchs2012multi}.

\item Support Vector Regression (SVR) trains two support regression models which are combined to predict the overall ratings \cite{jannach2012accuracy}.

\item  Criteria-Independent Contextual (CIC) model estimates multi-criteria ratings using a context-aware recommendation algorithm and subsequently applies SVR to predict overall ratings\cite{zheng2017criteria}.

\item  The Deep Multi-Criteria Collaborative Filtering (DMCF) utilizes a deep neural network to predict criteria ratings and learns the relationship between criteria and the overall rating \cite{nassar2020novel}.

\item  Deep Neural Network Matrix Factorization (DNN-MF) designed for information filtering in MCRS integrates respective models to capture non-linear interactions between users based on multi-criteria attributes \cite{sinha2022dnn}.

\item  Multi criteria based recommendations using Autoencoder and deep neural network with weight initialization by Firefly algorithm (MCAE-FADNN) employs deep autoencoders to predict missing criteria ratings and then builds non-linear interactions between users and items using a deep neural network  \cite{spoorthy2023multi}.

\item  Collective factor models (CFM) feature linear integration of loss functions for both overall ratings and multi-criteria ratings. The collective factor model is subsequently trained using both types of ratings, and they provided two types of predictions based on $CFM_{user}$ and $CFM_{item}$ \cite{fan2023improving}.
\end{enumerate}

\subsection{Parameter Sensitivity Analysis}

\begin{figure*}
  \centering
  \begin{subfigure}[b]{0.49\textwidth}
    \centering
    \includegraphics[width=\textwidth]{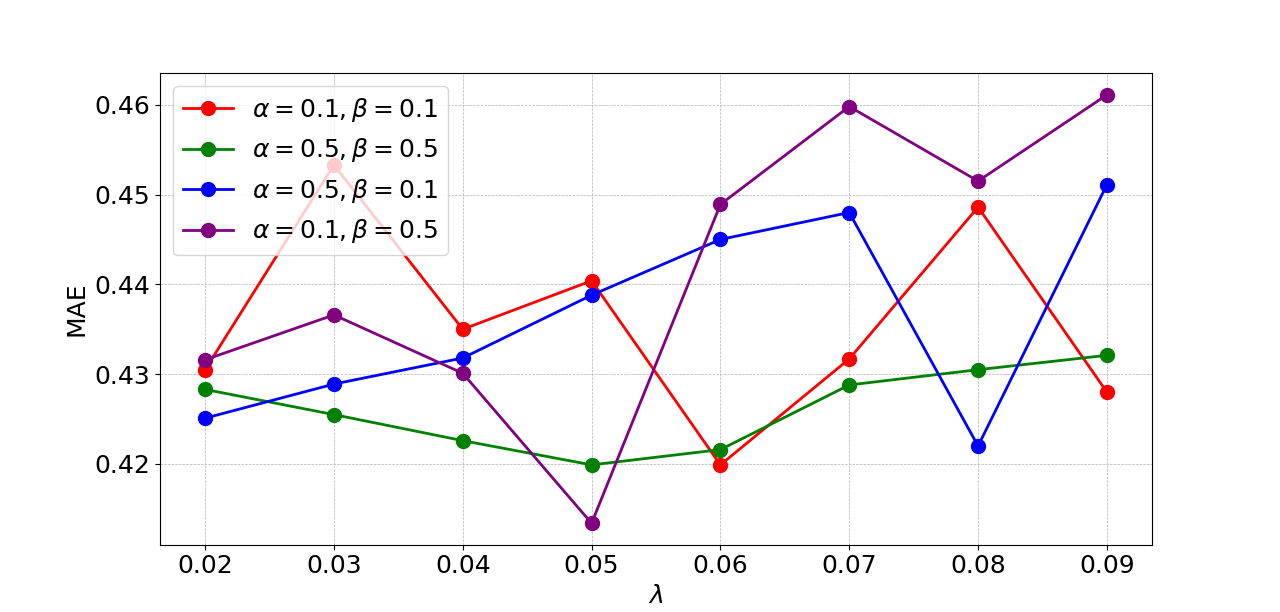}
    \caption{BeerAdvocate Dataset}
  \end{subfigure}
  \hfill 
  \hfill
  \begin{subfigure}[b]{0.49\textwidth}
    \centering
    \includegraphics[width=\textwidth]{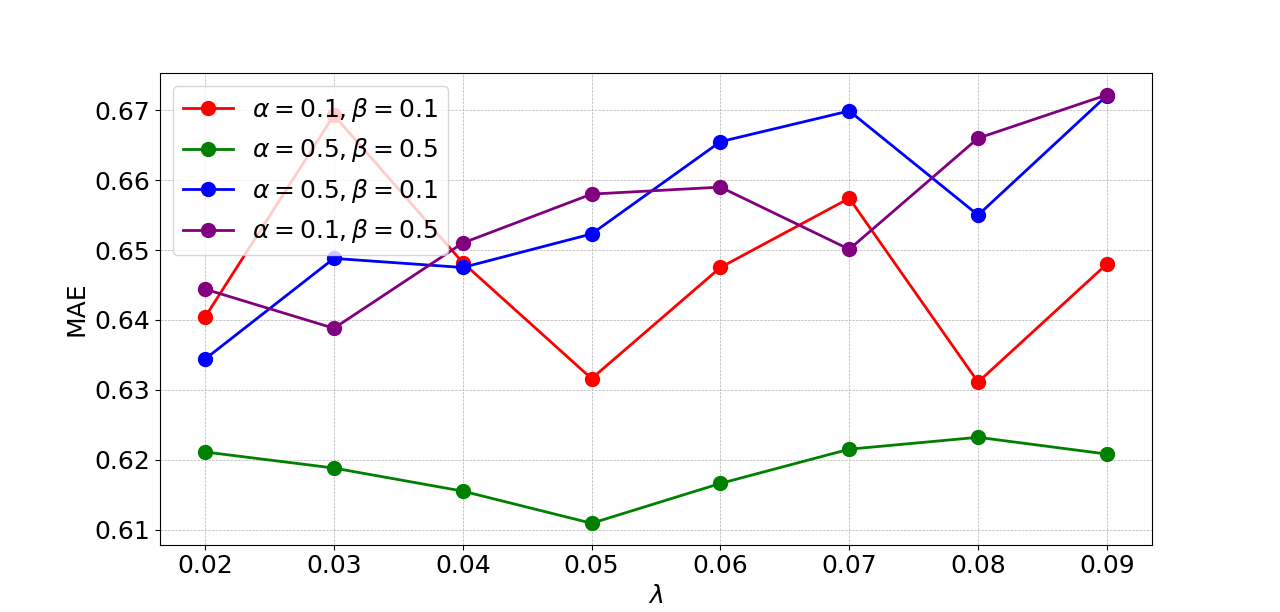}
    \caption{Yahoo!Movies Dataset}
  \end{subfigure}
  \hfill
  \caption{The effect of varying $\lambda$ values on MAE for the Yahoo!Movies and BeerAdvocate datasets, under different parameter settings for $\alpha$ and $\beta$. The parameters $\alpha$ and $\beta$ represent weights for balancing different components of the model, with four configurations explored: $\alpha = 0.1, \beta = 0.1$, $\alpha = 0.5, \beta = 0.5$, $\alpha = 0.5, \beta = 0.1$, and $\alpha = 0.1, \beta = 0.5$. Each curve shows how the MAE varies as $\lambda$, the regularization weight, changes across multiple values from 0.2 to 0.9.}
  \label{fig:Sensitivity}
\end{figure*}

We evaluate the performance of our D-MGAC framework by analyzing its hyperparameters, which influence the total loss function ($\mathcal{L}_{\text{total}}$). This loss function is composed of three key components: $\alpha$, which controls the weight of the mean absolute error (MAE) loss $\mathcal{L}_{P}$; $\beta$, which regulates the global weight similarity loss $\mathcal{L}_S$; and $\lambda$, which determines the influence of the $L2$ regularization term $|\mathbf{W}|_2^2$. To assess the sensitivity of these hyperparameters, we conducted experiments using various parameter configurations across two datasets (Yahoo!Movies and BeerAdvocate), as illustrated in Figure ~\ref{fig:Sensitivity}. We systematically explore different values of $\alpha$, $\beta$, and $\lambda$, with a particular emphasis on their impact on performance.

\begin{enumerate}
    \item \textbf{Equal Weighting of $\alpha$ and $\beta$}: Initially, we assign equal weight to both $\alpha$ and $\beta$ ($\alpha = 0.1, \beta = 0.1$ and $\alpha = 0.5, \beta = 0.5$). These configurations allow us to assess the impact of the $L2$ regularization parameter $\lambda$ while keeping $\alpha$ and $\beta$ constant.
    \item \textbf{Differentiated Weights for $\alpha$ and $\beta$}: In subsequent configurations, we vary the balance between $\alpha$ and $\beta$. Specifically, we explore the cases where $\alpha = 0.5, \beta = 0.1$ (higher priority for MSE loss) and $\alpha = 0.1, \beta = 0.5$ (higher priority for global weight similarity loss). These experiments help highlight the relative importance of these two terms in the loss function.
\end{enumerate}

For both datasets, we observe the sensitivity of the MAE to different values of $\lambda$. The results for Yahoo!Movies and BeerAdvocate datasets across different values of $\lambda$ (ranging from 0.2 to 0.9) for various $\alpha$ and $\beta$ combinations are plotted in Figure~\ref{fig:Sensitivity}. From these plots, it is evident that when $\alpha = 0.5$ and $\beta = 0.5$, we achieve more stable MAE across different values of $\lambda$, suggesting that balancing the two components leads to better performance. Increasing $\alpha$ or $\beta$ (i.e., prioritising either the MSE or global similarity loss) leads to variations in MAE, indicating the critical role of these hyperparameters in tuning model performance.

The optimal configuration of the hyperparameters is exemplified in Figure\ref{fig:Sensitivity}, where both $\alpha$ and $\beta$ exert the same influence, surpassing that of $\lambda$. The most favourable scenario occurs when the weights of $\alpha$ and $\beta$ are set to 0.5, while $\lambda$ is assigned a weight of 0.1. This observation suggests that the impact of local and global attention mechanisms is equitable, implying that both defined loss functions should carry equivalent weight in the optimization process. This balanced allocation of weights ensures a harmonious integration of local and global information in the model's decision-making process, thereby enhancing its overall performance.

\subsection{Embedding Dimension Sensitivity}

\begin{figure*}
  \centering
  \begin{subfigure}[b]{0.49\textwidth}
    \centering
    \includegraphics[width=\textwidth]{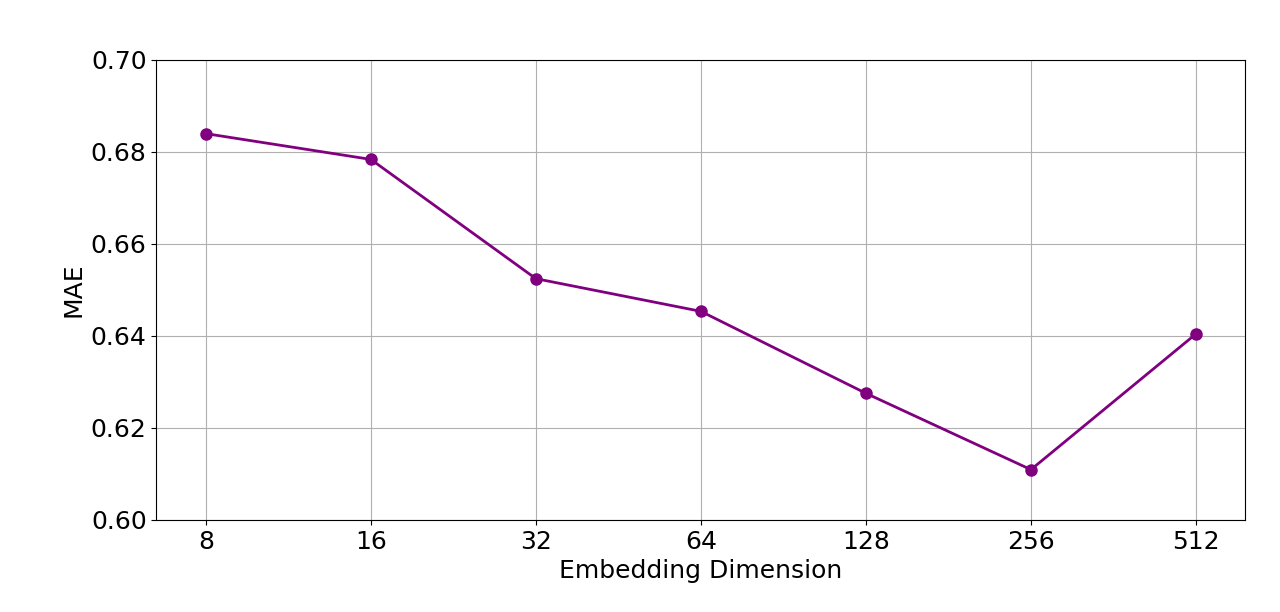}
    \caption{Yahoo!Movies}
  \end{subfigure}
  \hfill 
  \hfill
  \begin{subfigure}[b]{0.49\textwidth}
    \centering
    \includegraphics[width=\textwidth]{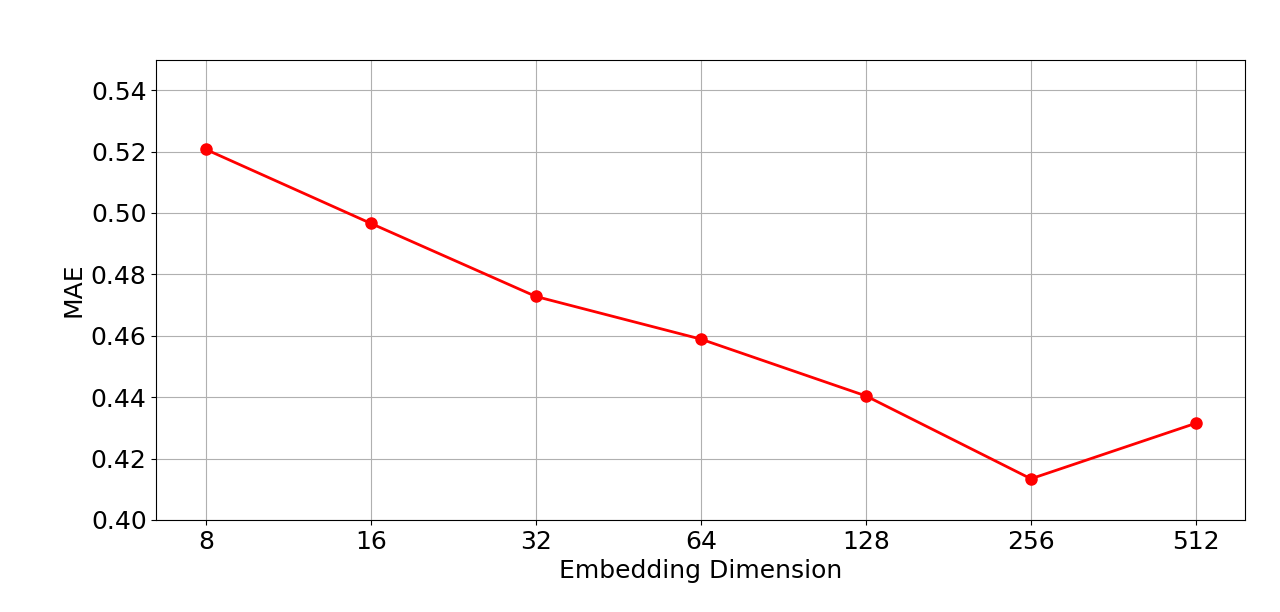}
    \caption{BeerAdvocate}
  \end{subfigure}
  \hfill
  \caption{Sensitivity analysis of the embedding dimension based on MAE}
  \label{fig:EmbDim}
\end{figure*}

We explore how varying embedding dimensions affect the sensitivity of recommendation accuracy generation based on the MAE and RMSE metrics. The experimental results are illustrated in Figure \ref{fig:EmbDim}.

The examination suggests that the accuracy of the recommendation increases as the dimensions of the encapsulation increase, reaching a peak at 256. This improvement comes from the ability of larger dimensions to encode information more effectively. However, excessive increases in dimensionality may result in a decrease in performance, potentially due to overfitting of the available data. Nonetheless, the performance of \textbf{D-MGAC} remains relatively steady across various embedding dimensions.

\subsection{Ablation study}

Next, we conduct an ablation study using an evaluation method based on our proposed approach. Such a study is crucial for isolating and understanding the contribution of each component in our model, allowing us to assess how specific design choices affect the overall performance. By systematically removing or altering parts of the model, we can identify which elements are essential for enhancing recommendation accuracy and which may be less critical or redundant. 

Our study includes three distinct experiments to explore different aspects of the model's performance. First, we investigate the impact of incorporating global attention into the results. This step is important as global attention is designed to capture long-range dependencies and interactions between nodes in the graph, and understanding its influence helps us verify its effectiveness in capturing critical relationships within the data. Next, we analyze the effects of excluding global contrastive learning from the loss function and compare the outcomes with and without this component. Global contrastive learning aims to improve the model's ability to distinguish between positive and negative samples by leveraging global structural information. By omitting it, we can determine whether this component is truly necessary for the model to learn better representations and make more accurate predictions. Finally, we explore how variations in the number of criteria for items influence the accuracy of our recommendation model. This experiment is key in determining how flexible and adaptable the model is to different levels of item complexity, as the number of criteria directly affects the granularity of the recommendation. Understanding this relationship helps optimize the model for diverse recommendation tasks with varying levels of detail. 

Each experiment in this ablation study is based on the MAE metric and includes different training sizes (Ts-40\%, Ts-60\%, Ts-80\%, and Ts-100\%). These experiments are designed to provide insights into epitomising the model's architecture and training process to enhance recommendation accuracy while also ensuring the model's interpretability and robustness.In our ablation study, we use three symbols to represent different versions of the proposed method. D-MGAC refers to the full model, while D-MGAC* and D-MGAC*- are defined as follows:

\begin{itemize}
    \item D-MGAC*: The model with global attention removed.
    \item D-MGAC*-: The model with both global attention and contrastive learning components removed.
\end{itemize}

\subsubsection{Impact of Global Attention}

The objective of this experiment is to evaluate the role and effectiveness of the global attention mechanism in our model. Global attention is designed to capture long-range dependencies and interactions between different nodes in the graph, which are essential to generate more accurate and personalized recommendations. By excluding this component, we aim to understand its contribution to the overall performance of the model.Our experimental setup involves comparing two conditions: one with global attention integrated and another without it. This allows us to assess whether global attention significantly enhances the model's ability to capture complex relationships in the data, such as latent connections between users and items across different contexts. Understanding this influence is important for determining whether the additional computational complexity introduced by global attention is justified in terms of improved recommendation quality. We analyze the results of these experiments on two datasets using the MAE metric, which includes the mean and standard deviation (Std = 30) over 30 runs, as presented in Table ~\ref{tab:Imp_GAtt}. This evaluation considers four different training sizes (Ts-40\%, Ts-60\%, Ts-80\%, and Ts-100\%) to provide a comprehensive understanding of how the absence of global attention affects model performance and recommendation outcomes. This comparison is critical for discerning the necessity of global attention and its potential trade-offs concerning model efficiency and accuracy. 
 
\begin{table*}
 \centering
 \small
 \caption{Evaluation of \textbf{D-MGAC} based on the impact of global attention, utilising the mean and standard deviation of the MAE from 30 independent model training runs while considering four different training sizes.}
 \label{tab:Imp_GAtt}
   \begin{tabular}{cccccccccc} 
     \toprule  
     \textbf{Dataset} & \textbf{Method} & \textbf{Ts-40\%} & \textbf{Ts-60\%} & \textbf{Ts-80\%} & \textbf{Ts-100\%} \\
     \midrule  
     \multirow{2}{*}{\textbf{Yahoo!Movies}} 
     & D-MGAC* & $0.6980_{(\pm 0.0145)}$  & $0.6733_{(\pm 0.008)}$  & $0.6566_{(\pm 0.0113)}$  & $0.6458_{(\pm 0.0048)}$  \\
     & D-MGAC & $0.6675_{(\pm 0.0091)}$  & $0.6412_{(\pm 0.0063)}$  & $0.6257_{(\pm 0.0084)}$ & $0.6105_{(\pm 0.0058)}$      \\
     \midrule
     \multirow{2}{*}{\textbf{BeerAdvocate}} 
     & D-MGAC* & $0.5122_{(\pm 0.0352)}$  & $0.4966_{(\pm 0.0115)}$  & $0.4670_{(\pm 0.0064)}$  & $0.4752_{(\pm 0.0057)}$  \\
     & D-MGAC & $0.4774_{(\pm 0.0079)}$  & $0.4608_{(\pm 0.0086)}$  & $0.4314_{(\pm 0.0072)}$ & $0.4156_{(\pm 0.0036)}$  \\
     \bottomrule
   \end{tabular}
\end{table*}

The ablation study in Table \ref{tab:Imp_GAtt} evaluates the performance of the D-MGAC model compared to a method that lacks global attention in two datasets, using MAE as the metric. 

The results reveal a clear performance gap between the complete model (D-MGAC) and the version without global attention (D-MGAC*) across both datasets and all training sizes. For the Yahoo!Movies dataset, the MAE consistently improves when global attention is included. For instance, at Ts-100\%, the MAE for D-MGAC is 0.6105, whereas it is 0.6458 for D-MGAC*, indicating that global attention reduces the prediction error. Similar trends are observed for the other training sizes, confirming that global attention enhances the model’s ability to capture complex, long-range dependencies between users and items. In the BeerAdvocate dataset, the impact of global attention is even more pronounced. At Ts-100\%, D-MGAC achieves an MAE of 0.4156, significantly outperforming the D-MGAC* model, which has an MAE of 0.4752. This larger performance gap suggests that global attention is particularly beneficial when dealing with datasets characterised by rich user-item interactions and varied preferences, as seen in BeerAdvocate. Across both datasets, the MAE results show lower variance (i.e., standard deviation) when global attention is included, indicating that the model is not only more accurate but also more stable across different training runs. This further supports the hypothesis that global attention improves the model's robustness by providing a more comprehensive understanding of user-item relationships. Overall, the inclusion of global attention in D-MGAC leads to more precise and consistent recommendations, with notable improvements in both accuracy and stability, justifying its computational cost in terms of enhanced model performance.

\subsubsection{Impact of Global Attention and Contrastive Learning}

We investigate the impact of excluding the global attention mechanism and the global contrastive learning term from our loss function, both of which are crucial for enhancing the model's discriminative capabilities by pulling similar items closer within the latent space. Through an ablation study, we compare the performance of our recommendation model under two scenarios: one where the complete proposed model (D-MGAC) is used, and another where the global attention and global contrastive learning components are omitted (D-MGAC*-). This analysis aims to determine how the absence of global attention and contrastive learning affects the model's ability to differentiate between items and make accurate recommendations. Our experimental evaluation spans two distinct datasets, allowing us to observe how the removal of these components influences the precision and robustness of the recommendations.

\begin{table*}
 \centering
 \small
 \caption{Evaluation of \textbf{D-MGAC} based on the impact of global attention and contrastive learning, utilising the mean and standard deviation (in brackets) of the MAE from 30 independent model training runs across four different training sizes. The results are reported on a fixed test dataset.}
 \label{tab:Ab_GCL}
   \begin{tabular}{cccccccccc} 
   \toprule  
   \textbf{Dataset} & \textbf{Method} & \textbf{Ts-40\%} & \textbf{Ts-60\%} & \textbf{Ts-80\%} & \textbf{Ts-100\%} \\
   \midrule   
   \multirow{2}{*}{\textbf{Yahoo!Movies}} 
   & D-MGAC*- & $0.7248_{(\pm 0.0244)}$  & $0.6912_{(\pm 0.012)}$  & $0.6732_{(\pm 0.0572)}$  & $0.6617_{(\pm 0.0074)}$  \\
   & D-MGAC & $0.6675_{(\pm 0.0091)}$  & $0.6412_{(\pm 0.0063)}$  & $0.6257_{(\pm 0.0084)}$ & $0.6105_{(\pm 0.0058)}$      \\
   \midrule
   \multirow{2}{*}{\textbf{BeerAdvocate}}
   & D-MGAC*- & $0.5429_{(\pm 0.0152)}$  & $0.5178_{(\pm 0.0088)}$  & $0.5061_{(\pm 0.0084)}$  & $0.4903_{(\pm 0.0064)}$   \\
   & D-MGAC & $0.4774_{(\pm 0.0079)}$  & $0.4608_{(\pm 0.0086)}$  & $0.4314_{(\pm 0.0072)}$ & $0.4156_{(\pm 0.0036)}$  \\
   \bottomrule
   \end{tabular}
\end{table*}

The results in Table \ref{tab:Ab_GCL} clearly illustrate the significant impact of both global attention and global contrastive learning on the performance of the D-MGAC model. For the Yahoo!Movies dataset, the model that excludes both global attention and contrastive learning (D-MGAC*-) exhibits a markedly higher MAE across all training sizes. For instance, at Ts-100\%, the MAE for D-MGAC*- is 0.6617, compared to 0.6105 for D-MGAC, indicating that the complete model significantly reduces prediction error. The trend persists across the other training sizes, with MAE values for D-MGAC consistently lower than those of D-MGAC*- (e.g., at Ts-40\%, D-MGAC has an MAE of 0.6675, while D-MGAC*- shows 0.7248). This suggests that the absence of global attention and contrastive learning adversely affects the model’s ability to discern relationships and provide accurate recommendations. Similarly, in the BeerAdvocate dataset, the impact is pronounced. The D-MGAC*- model shows a higher MAE at all training sizes, with the most significant gap appearing at Ts-100\%, where D-MGAC achieves an MAE of 0.4156 while D-MGAC*- stands at 0.4903. This consistent under performance of D-MGAC* across both datasets highlights the importance of integrating both global attention and contrastive learning to enhance the model’s discriminative power. The standard deviations presented in the results indicate that the model with the complete setup (D-MGAC) not only achieves lower MAE values but also demonstrates improved stability and consistency in its predictions. For example, the standard deviation for D-MGAC at Ts-100\% is 0.0058, which is smaller than that of D-MGAC* (0.0074), suggesting that the full model is more reliable across different runs. Overall, the ablation study clearly demonstrates that both global attention and contrastive learning are integral to the D-MGAC model's effectiveness. The removal of these components leads to significantly higher MAE values, underscoring their crucial roles in enhancing the model's ability to make accurate and robust recommendations across different datasets.

\subsubsection{Impact of the Number of Criteria}

We conducted an ablation study to explore how changes in the number of criteria impact the performance of D-MGAC where we incrementally increased the criteria from one to four for Yahoo!Movies and BeerAdvocate. This setup enabled us to observe how the model's performance evolves with the addition of each criterion. Table ~\ref{tab:Criteria} provides information on how the criteria affect the precision of the recommendation model on various data sets using MAE and RMSE. 
It is evident that as the number of criteria increases, there is generally an improvement in the accuracy of the metrics that is observable across the different datasets and metrics. For example, considering Yahoo!Movies, as the number of criteria increases from 1 to 4, there is a decrease in both MAE and RMSE, indicating improved accuracy. Similarly, in the BeerAdvocate dataset, there is a reduction in MAE and RMSE as the number of criteria increases.

\begin{table*}
  \centering
  \caption{ MAE of the \textbf{D-MGAC} based on the number of criteria utilising the mean and standard deviation (in brackets) from 30 experimental runs conducted on the test dataset.}
  \label{tab:Criteria}
    \begin{tabular}{ccccc} 
      \toprule  
      \textbf{Dataset} & \textbf{1 criterion} & \textbf{2 criteria} & \textbf{3 criteria} & \textbf{4 criteria} \\ 
      \midrule  
      \multirow{1}{*}{\textbf{Yahoo!Movies}}  & $0.7316_{(\pm 0.0175)}$ & $0.6870_{(\pm 0.0126)}$ & $0.6404_{(\pm 0.0079)}$ & $\mathbf{0.6105_{(\pm 0.0058)}}$    \\
      \midrule
      \multirow{1}{*}{\textbf{BeerAdvocate}}  & $0.5264_{(\pm 0.0075)}$ & $0.4720_{(\pm 0.0079)}$ & $0.4453_{(\pm 0.0054)}$ & $\mathbf{0.4156_{(\pm 0.0036)}}$  \\
      \bottomrule
    \end{tabular}
\end{table*}

The analysis of Table \ref{tab:Criteria} reveals a significant correlation between the number of criteria utilized in the D-MGAC model and its performance, as evidenced by the MAE results across the Yahoo!Movies and BeerAdvocate datasets. In the Yahoo!Movies dataset, the MAE decreases from 0.7316 with one criterion to 0.6105 when four criteria are employed, indicating a marked improvement in predictive accuracy. This trend reflects the model's enhanced capability to capture user preferences and item characteristics as more criteria are introduced, with the relatively low standard deviations suggesting consistent performance across the 30 experimental runs. Similarly, the BeerAdvocate dataset demonstrates a decrease in MAE from 0.5264 for one criterion to 0.4156 for four criteria, further corroborating the benefits of multiple criteria in improving model accuracy. The standard deviations here also remain low, reinforcing the reliability of the model’s predictions. Overall, the findings underscore the importance of incorporating multiple criteria in recommendation systems, as the D-MGAC model exhibits enhanced performance and stability, allowing for more precise item differentiation and recommendations, thereby enhancing user satisfaction across diverse datasets.

\subsection{D-MGAC on the TripAdvisor}

\subsubsection{Evaluation of D-MGAC on the TripAdvisor Dataset}

In this section, we evaluate the various components of the D-MGAC model using the TripAdvisor dataset to assess its performance across different configurations. Specifically, we focus on comparing D-MGAC, D-MGAC*, and D-MGAC*- under varying training data segments (TS), which are defined by the percentage of the dataset used for training. For example, TS-40 refers to a training set where only 40\% of the data is randomly selected for each experimental run, while the remaining 60\% is excluded from training. The test data, however, remains constant across all experiments and is not altered. The performance of the models is evaluated based on MAE, and the results are presented in Table ~\ref{tab:Ab_Trip}. The table highlights the mean MAE and standard deviation across 30 independent model runs for each training segment. By varying the amount of training data, we aim to determine how well each model generalises to unseen data. This analysis allows us to explore how the models handle varying amounts of data and to compare their effectiveness under different training scenarios. The comparison also serves to evaluate the robustness and stability of the models. The standard deviation provides insights into the consistency of the models' performance, indicating how much variance there is between different runs with different data samples. Understanding this behavior is crucial, especially when working with real-world, noisy datasets like TripAdvisor, where the availability of complete data is often a challenge.

\begin{table*}
    \centering
    \caption{ Comparison of D-MGAC*, D-MGAC*-, and D-MGAC for different sets of training data, defined by training segments (TS). We report the mean MAE and standard deviation (in brackets) across 30 independent model training runs.}
    \label{tab:Ab_Trip}
        \begin{tabular}{lcccccc}  
            \toprule
            \textbf{Dataset} & \textbf{Method} & \textbf{Phase} & \textbf{TS-40\%} & \textbf{TS-60\%} & \textbf{TS-80\%} & \textbf{TS-100\%} \\
            \midrule
            \multirow{6}{*}{TripAdvisor} 
            & D-MGAC* & Train & $0.8053_{(\pm 0.004)}$  & $0.7838_{(\pm 0.0031)}$ & $0.7541_{(\pm 0.0024)}$ & $0.7461_{(\pm 0.0032)}$ \\  
            &  & Test & $0.8294_{(\pm 0.047)}$  & $0.8006_{(\pm 0.029)}$ & $0.7726_{(\pm 0.039)}$ & $0.7532_{(\pm 0.0235)}$ \\
            \cmidrule{3-7}
            
            & D-MGAC*- & Train & $0.8345_{(\pm 0.0062)}$  & $0.8107_{(\pm 0.0055)}$ & $0.7647_{(\pm 0.0073)}$ & $0.7583_{(\pm 0.003)}$ \\  
            &  & Test & $0.8517_{(\pm 0.0462)}$  & $0.8217_{(\pm 0.0351)}$ & $0.7954_{(\pm 0.038)}$ & $0.7704_{(\pm 0.0156)}$ \\
            \cmidrule{3-7}
            
            & D-MGAC  & Train & \bf{$0.7608_{(\pm 0.0036)}$} & \bf{$0.7427_{(\pm 0.0041)}$} & \bf{$0.7315_{(\pm 0.0035)}$} & \bf{$0.7168_{(\pm 0.0028)}$} \\
            &  & Test & \bf{$0.7712_{(\pm 0.0265)}$} & \bf{$0.7528_{(\pm 0.0186)}$} & \bf{$0.7387_{(\pm 0.0213)}$} & \bf{$0.7274_{(\pm 0.0106)}$} \\
            \bottomrule
        \end{tabular}
\end{table*}

The results in Table \ref{tab:Ab_Trip} reveal the performance differences between D-MGAC, D-MGAC*, and D-MGAC*- across various training segments (TS-40\%, TS-60\%, TS-80\%, TS-100\%). The full D-MGAC model consistently outperforms the variants, demonstrating the lowest MAE in both training and testing phases across all training data splits. Specifically, D-MGAC achieves the best test MAE of 0.7274 with 100\% training data, indicating the model's superior generalization capability when both global attention and contrastive learning components are present. In contrast, D-MGAC*, which removes global attention, shows higher MAE, highlighting the importance of attention in capturing key relationships. D-MGAC*-, which lacks both global attention and contrastive learning, performs the worst, especially at smaller training segments like TS-40\%, with a test MAE of 0.8294, emphasising the cumulative benefit of these components for improving model accuracy as more data becomes available. Overall, the inclusion of both global attention and contrastive learning in D-MGAC leads to more accurate and stable predictions.

In the following, we evaluate the performance of the proposed D-MGAC method on the TripAdvisor dataset, focusing on the impact of varying the number of criteria. The evaluation is based on the MAE metric, with the results averaged over 30 experimental runs, including the standard deviation to indicate consistency. Table \ref{tab:Trip_Criteria} presents the results, highlighting the relationship between the number of criteria used in the model and its performance, both in terms of training and testing.

\begin{table*}
  \centering
  \caption{  Evaluation of the \textbf{D-MGAC} based on the number of criteria. This includes the mean and standard deviation (in brackets) over 30 experimental runs on the TripAdvisor dataset, showing Train and Test of MAE.}
  \label{tab:Trip_Criteria}
    \begin{tabular}{cccc} 
      \toprule  
      \textbf{Database} & \textbf{Criteria} & \textbf{Train MAE} & \textbf{Test MAE} \\ 
      \midrule  
      \multirow{7}{*}{TripAdvisor Dataset} 
      & One criterion   & $0.8254_{(\pm 0.0048)}$ & $0.8415_{(\pm 0.0258)}$ \\
      & Two criteria    & $0.7732_{(\pm 0.0029)}$ & $0.7874_{(\pm 0.0164)}$ \\
      & Three criteria  & $0.7534_{(\pm 0.0038)}$ & $0.7612_{(\pm 0.0271)}$ \\
      & Four criteria   & $0.7426_{(\pm 0.0042)}$ & $0.7488_{(\pm 0.0247)}$ \\
      & Five criteria   & $0.7350_{(\pm 0.0030)}$ & $0.7410_{(\pm 0.0155)}$ \\
      & Six criteria    & $0.7302_{(\pm 0.0033)}$ & $0.7366_{(\pm 0.0183)}$ \\
      & Seven criteria  & $0.7168_{(\pm 0.0028)}$ & $0.7274_{(\pm 0.0106)}$ \\
      \bottomrule
    \end{tabular}
\end{table*}

The results in Table \ref{tab:Trip_Criteria} show that as the number of criteria increases, the D-MGAC model consistently improves its performance, with both training and test MAE steadily decreasing. Training MAE drops from 0.8254 for 1 criterion to 0.7168 for 7 criteria, reflecting enhanced error minimisation due to richer input features. Similarly, test MAE decreases from 0.8415 to 0.7274, indicating better generalisation to unseen data. The low standard deviations across both training and testing suggest stable and reliable model behavior across runs. Overall, the results demonstrate that using more criteria enhances the model’s predictive accuracy, though computational costs may rise with additional criteria.

\subsection{Baseline methods vs D-MGAC }

\begin{table*}
  \caption{ Comparison of D-MGAC with other methods based on MAE for 4-criteria in MCRS datasets. We present the best results from previous methods, along with the mean and standard deviation (in brackets) from 30 experimental runs for D-MGAC.}
  \centering
    \begin{tabular}{llcccccc}
      \toprule
      \textbf{ID} & \textbf{Algorithm} & \multicolumn{2}{c}{Yahoo!Movies} & \multicolumn{2}{c}{BeerAdvocate} \\
      \cmidrule(lr){3-4} \cmidrule(lr){5-6}
      & & \textbf{MAE} & \textbf{RMSE} & \textbf{MAE} & \textbf{RMSE} \\
      \midrule
      1 & BMF \cite{paterek2007improving} & 0.6289 & 0.8646 & 0.4394 & 0.5858 \\
      2 & MSVD \cite{li2008improving} & 0.6332 & 0.8738 & 0.4473 & 0.5960 \\
      3 & UserKNN \cite{peterson2009k} & 0.9260 & 1.2329 & 0.6559 & 0.8444 \\
      4 & MLR \cite{fuchs2012multi} & 0.6326 & 0.8664 & 0.4442 & 0.5929 \\
      5 & SVR \cite{jannach2012accuracy} & 0.6248 & 0.8671 & 0.4470 & 0.5993 \\
      6 & CIC \cite{zheng2017criteria} & 0.6200 & 0.8782 & 0.4429 & 0.5914 \\
      7 & DMCF \cite{nassar2020novel} & 0.7012 & 0.9139 & 0.4698 & 0.6240 \\
      8 & DNN-MF \cite{sinha2022dnn} & 0.6178 & 0.8606 & 0.4483 & 0.6077 \\
      9 & MCAE-FADNN \cite{spoorthy2023multi} & 0.6277 & 0.8793 & 0.4698 & 0.6240 \\
      10 & MultiUserKNN \cite{fan2023improving} & 0.9319 & 1.2396 & 0.6572 & 0.8441 \\
      11 & $CFM_{user}$ \cite{fan2023improving} & 0.6184 & 0.8802 & 0.4403 & 0.5904 \\
      12 & $CFM_{item}$ \cite{fan2023improving} & 0.6127 & 0.8433 & 0.4408 & 0.5904 \\
      13 & \textbf{D-MGAC} & $\textbf{0.6105}_{(\pm 0.0058)}$ & $\textbf{0.7219}_{(\pm 0.0141)}$ & $\textbf{0.4156}_{(\pm 0.0036)}$ & $\textbf{0.5197}_{(\pm 0.0059)}$ \\
      \bottomrule
    \end{tabular}
  \label{tab:Evaluation}
\end{table*}

In Table \ref{tab:Evaluation}, we compare the performance of our D-MGAC framework against a range of baseline methods using MAE and RMSE metrics for the Yahoo!Movies and BeerAdvocate datasets. Although D-MGAC achieves the best overall performance in both datasets, it is important to note that some baseline methods produce results that are quite competitive. For example, $CFM_{item}$ closely matches the performance of D-MGAC in both MAE and RMSE on the Yahoo!Movies dataset, with only a slight difference in the final RMSE score (0.7219 vs. 0.8433). Similarly, DNN-MF, a deep learning-based matrix factorization approach, delivers an RMSE that is very close to D-MGAC (0.8606 vs. 0.7219), suggesting that its architecture effectively captures user-item interactions, though it lacks the multi-relational attention mechanisms of D-MGAC.

On the BeerAdvocate dataset, $CFM_{user}$ and $CFM_{item}$ also come close to D-MGAC, particularly in terms of RMSE, with values of 0.5904 compared to D-MGAC's 0.5628. This indicates that both methods are well-suited for handling multi-criteria recommendation tasks. However, D-MGAC's superior performance can be attributed to its use of multi-relational attention and local-global integration, which enables it to better capture the nuances of user preferences and complex item relationships across multiple criteria. While D-MGAC consistently delivers the best performance, these close results highlight that some baseline methods are well-optimized for the datasets at hand, and their competitiveness suggests that the choice of model may depend on specific use cases and computational considerations. Nevertheless, D-MGAC stands out due to its ability to simultaneously model multi-relational interactions and capture fine-grained user preferences, contributing to its improved MAE and RMSE across both datasets.

\section{Discussion}

In this study, we presented the D-MGAC framework for MCRS that leverages advanced embedding techniques utilising bipartite graphs to represent users and items across different domains, with multiple weighted edges capturing various criteria. Each edge signifies a rating for a specific criterion. Through multiviews, the relationship between users and items based on each criterion is converted into one view, with each view displaying the relationship between users and items based on a single criterion. In the subsequent stage, our model employs both local and global attention mechanisms to consider node relationships within each view as well as across the entire graph. Our objective is to enhance node representation learning in recommender systems by integrating MDGAT \cite{yuan2023cross}, which encompasses local and global attention stages. These attention mechanisms play a pivotal role in effectively capturing node interactions. The model consists of two graph attention convolutional layers aimed at capturing crucial node relationships, thus addressing challenges such as detecting hidden correlations and capturing indirect dependencies.

Traditionally, attention mechanisms have been applied independently to each view, potentially missing intricate relationships \cite {xie2020mgat}. To overcome this, we propose a novel approach using L-BGNN adjacency matrices for each criterion, capturing relationship strengths between nodes based on specific criteria. This method enables the discovery of popular criteria for each item, preserving critical relationships within each view. During training, formulating the loss function is crucial. Our model processes multiview data represented by bipartite graphs and utilises dual attention to generate embedding vectors. 

Establishing connections between nodes within view-based embeddings poses a challenge. We use edge weights between users and items in each view to address this issue, assessing user interest using specific criteria. However, in multiview networks, each view typically represents a distinct and often sparse and biased relationship among nodes \cite{he2020game, cui2021mvgan}. Overcoming this limitation involves integrating information from other viewpoints to gain a comprehensive understanding. Our approach addresses this challenge by evaluating node similarity across multiple perspectives using embedding vectors and a cost function with learnable attention weight parameters. This analysis helps identify nodes that demonstrate similarity across views, align node embeddings, and enhance coherence. The attention weights control the significance of each view in computing the similarity loss, and the model dynamically adjusts them during training for improved performance.

Lastly, the proposed model underwent evaluation using two real datasets and 12 basic algorithms. By comparing the results using two metrics, MAE and RMSE, the efficiency of the proposed method was demonstrated in contrast to other methods analysed. These innovations empower our model to simultaneously capture fine-grained, view-specific details and overarching global patterns, significantly improving its effectiveness in MCRS. The model offers enhanced personalisation through dual-attention and contrastive learning strategies, overcoming previous limitations and introducing innovative loss functions for effective training.

One major limitation of our study is the absence of an in-depth analysis of the model's behaviour in real-world scenarios, particularly under conditions of high variability in user preferences and item availability. The two datasets used for evaluation may not fully represent the diversity of recommendation scenarios encountered in practice. Additionally, while the comparison with 12 baseline algorithms demonstrates the competitive performance of our approach, it does not provide insights into how the model behaves in dynamic or cold-start scenarios, where user-item interactions are limited or nonexistent. Another important consideration is the model's sensitivity to hyperparameter tuning, particularly in the attention weights used to control the significance of each view. Improper tuning of these parameters could lead to overemphasis on certain criteria, potentially skewing the recommendations toward specific aspects of user behavior while neglecting others. Additionally, while the use of attention mechanisms helps mitigate some of the biases present in sparse and biased views, it does not entirely eliminate them.

In terms of future directions, there are several avenues for improvement. First, integrating dynamic hypergraph learning techniques into the D-MGAC framework could address some of the limitations related to static adjacency matrices and evolving user preferences. Hypergraph-based methods would allow us to model more complex relationships between users and items, capturing higher-order interactions that go beyond pairwise relationships. This could be particularly useful in addressing sparsity and improving the model's ability to generalise across different criteria. Furthermore, exploring advanced optimization techniques, such as adaptive hypergraph pruning and regularization strategies, could help reduce the model's computational complexity without sacrificing performance. Dynamic weighting mechanisms could also be introduced to adjust the contribution of each view based on context, user behavior, or item characteristics, providing a more personalized and context-aware recommendation experience.

\section{Conclusion}

This study introduced the D-MGAC framework for MCRS, combining dual-attention mechanisms with contrastive learning to address challenges such as data sparsity and bias in multiview networks. Using both local and global attention, the model captures nuanced relationships between users and items, offering enhanced personalization and improved recommendation accuracy on different criteria. However, the framework is not without limitations. The complexity introduced by dual attention mechanisms and contrastive learning increases computational overhead, potentially impacting scalability for large-scale datasets. Additionally, the model's reliance on fixed adjacency matrices may struggle to adapt to dynamic user preferences over time. Moreover, while attention mechanisms help to manage the sparsity, they may not fully eliminate biases present in certain views. The key takeaway from our work is that D-MGAC significantly advances the capability of MCRS by effectively integrating multiview information, but future work must address its limitations, particularly in scalability and adaptability. In summary, while the model offers substantial improvements in recommendation accuracy, further optimization is necessary to ensure its broader applicability and robustness in diverse real-world scenarios.

In conclusion, while the D-MGAC framework demonstrates significant potential in improving recommendation accuracy through multiview attention mechanisms, its limitations in scalability, handling dynamic preferences, and sensitivity to hyperparameters should not be overlooked. Addressing these challenges will be critical to ensuring the model's applicability to a broader range of recommendation scenarios and enhancing its robustness in real-world environments.

\section{Data and Code availability}

Data and Python code for the D-MGAC model are provided via the GitHub repository:

\url{https://github.com/sydney-machine-learning/multicriteria-recommendersys}.

\bibliographystyle{elsarticle-num}
\bibliography{refs.bib}

\appendix
\section{\centering Comparison of Various Methodologies Using Single and Multi-Criteria Recommender Systems} \label{app:comparison}
\small
\begin{tabular}{lp{.25\textwidth}}
\hline
Model & Methodology\\
\hline
\multicolumn{2}{c}{\textbf{\emph{Single-Criterion}}}\\
\textbf{GRASER} \cite{gwadabe2022improving} & Attention and MLP\\
\textbf{DiffNet++} \cite{wu2020diffnet++} & Graph Neural Networks \\
\textbf{NHybF} \cite{berkani2022neural} & Deep Learning and MLP\\
\textbf{MVL} \cite{santosh2020mvl} &  Multiview Learning and Attention\\
\textbf{NGCF} \cite{wang2019neural} & Graph Neural Networks \\
\textbf{HRDR} \cite{liu2020hybrid} & Convolutional  Neural Networks and MLP \\
\textbf{ELECTRA} \cite{gheewala2024exploiting} & Deep Learning and MLP \\
\textbf{CNNMF} \cite{hien2024deep} & Deep Learning \\
\textbf{CoSFL-MVC} \cite{yan2023collaborative} & Multiview Learning \\
\textbf{AAIN} \cite{he2023aain} & Deep Learning and Multiview Learning \\
\textbf{KFAtt} \cite{liu2020kalman} & Deep Learning and Attention \\
\textbf{IFLM} \cite{zhang2022embedding} & Graph Neural Networks \\
\textbf{MV-GAN} \cite{chen2022multi} & Multiview Learning and Attention \\
\textbf{WAUC} \cite{lin2024mixed} & Attention and MLP\\
\textbf{MSMT-LSI} \cite{wang2024sequential} & Attention \\
\textbf{Mmkdgat} \cite{wang2024mmkdgat} & Multiview Learning and Attention \\
\textbf{MICL} \cite{li2024multi} &  Multiview Learning \\
\textbf{KMVG} \cite{chen2023knowledge} & Attention \\
\hline
\multicolumn{2}{c}{\textbf{\emph{Multi-Criteria}}}\\
\textbf{UserKNN} \cite{fan2023improving} & Neighborhood-Based Collaborative Filtering \\
\textbf{MultiUserKNN} \cite{fan2023improving} & Neighborhood-Based Collaborative Filtering \\
\textbf{BMF} \cite{paterek2007improving} & Matrix Factorization \\
\textbf{MSVD} \cite{li2008improving} & Decomposition-based method \\
\textbf{MLR} \cite{fuchs2012multi} & Regression-based method \\
\textbf{SVR} \cite{jannach2012accuracy} & Regression-based method \\
\textbf{CIC} \cite{zheng2017criteria} & Criteria-Independent Contextual \\
\textbf{DMCF} \cite{nassar2020novel} & Deep Neural Networks \\
\textbf{DMCF+} \cite{nassar2020multi} & Deep Neural Networks \\
\textbf{AEMC} \cite{shambour2021deep} & Deep Learning \\
\textbf{MCTc–C2R} \cite{hong2021multi} & Tensor factorization \\
\textbf{SVR-SR} \cite{zhang2021multi} & Matrix factorization and clustering  \\
\textbf{P-DNN-MCCF} \cite{rismala2024personalized} & Neural networks \\
\textbf{DeepMF} \cite{singh2024comparative} & Deep MF and MLP \\
\textbf{DNN-MF} \cite{sinha2022dnn} & Graph Neural Networks  \\
\textbf{MCAE-FADNN} \cite{spoorthy2023multi} & Graph Neural Networks  \\
$\textbf{CFM}_{\text{user}}$ \cite{fan2023improving} & Collective Factor Model \\
$\textbf{CFM}_{\text{item}}$ \cite{fan2023improving} & Collective Factor Model \\
\hline
\end{tabular}


\end{document}